\documentclass[twocolumn]{aastex63}

\usepackage{url}

\usepackage{gensymb}
\usepackage{amsmath}
\usepackage{flafter}
\usepackage{multirow} 
\usepackage{xcolor}

\graphicspath{{./}{figures/}}
\newcommand{\hx}{\em HXMT}
\newcommand{\Nu}{\em NuSTAR}
\newcommand{\sw}{\em Swift}
\newcommand{\exo}{EXO~1846--031}

\shorttitle{A variable ionized disk wind in EXO~1846--031}

\begin{document}

\title{A variable ionized disk wind in the black-hole candidate {\exo}}

\author{Yanan Wang}
\affil{Universit\'e de Strasbourg, CNRS, Observatoire astronomique de Strasbourg, UMR 7550, F-67000 Strasbourg, France}
\altaffiliation{E-mail: yanan.wang@astro.unistra.fr}
\affil{Physics \& Astronomy, University of Southampton, Southampton, Hampshire SO17~1BJ, UK}

\author{Long Ji} 
\affil{Institut für Astronomie und Astrophysik, Kepler Center for Astro and Particle Physics, Eberhard Karls Universität, Sand 1, D-72076 Tübingen, Germany}

\author{Javier A. Garc{\'\i}a} 
\affil{Cahill Center for Astronomy and Astrophysics, California Institute of Technology, Pasadena, CA 91125, USA}
\affil{Dr Karl Remeis-Observatory and Erlangen Centre for Astroparticle Physics, Sternwartstr. 7, D-96049 Bamberg, Germany}

\author{Thomas Dauser} 
\affil{Dr Karl Remeis-Observatory and Erlangen Centre for Astroparticle Physics, Sternwartstr. 7, D-96049 Bamberg, Germany}

\author{Mariano M\'endez} 
\affil{Kapteyn Astronomical Institute, University of Groningen, PO Box 800, 9700 AV Groningen, The Netherlands}

\author{Junjie Mao}
\affil{Department of Physics, University of Strathclyde, Glasgow G4 0NG, UK}

\author{L. Tao}
\affil{Key Laboratory for Particle Astrophysics, Institute of High Energy Physics, Chinese Academy of Sciences, 19B Yuquan Road, Beijing 100049, China}

\author{Diego Altamirano} 
\affil{Physics \& Astronomy, University of Southampton, Southampton, Hampshire SO17~1BJ, UK}

\author{Pierre Maggi} 
\affil{Universit\'e de Strasbourg, CNRS, Observatoire astronomique de Strasbourg, UMR 7550, F-67000 Strasbourg, France}

\author{S.~N. Zhang}
\affil{Key Laboratory for Particle Astrophysics, Institute of High Energy Physics, Chinese Academy of Sciences, 19B Yuquan Road, Beijing 100049, China}
\affil{University of Chinese Academy of Sciences, Chinese Academy of Sciences, Beijing 100049, China}

\author{M. Y. Ge}
\affil{Key Laboratory for Particle Astrophysics, Institute of High Energy Physics, Chinese Academy of Sciences, 19B Yuquan Road, Beijing 100049, China}

\author{L. Zhang} 
\affil{Physics \& Astronomy, University of Southampton, Southampton, Hampshire SO17~1BJ, UK}

\author{J.~L. Qu}
\affil{Key Laboratory for Particle Astrophysics, Institute of High Energy Physics, Chinese Academy of Sciences, 19B Yuquan Road, Beijing 100049, China}
\affil{University of Chinese Academy of Sciences, Chinese Academy of Sciences, Beijing 100049, China}

\author{S. Zhang}
\affil{Key Laboratory for Particle Astrophysics, Institute of High Energy Physics, Chinese Academy of Sciences, 19B Yuquan Road, Beijing 100049, China}

\author{X. Ma}
\affil{Key Laboratory for Particle Astrophysics, Institute of High Energy Physics, Chinese Academy of Sciences, 19B Yuquan Road, Beijing 100049, China}

\author{F. J. Lu}
\affil{Key Laboratory for Particle Astrophysics, Institute of High Energy Physics, Chinese Academy of Sciences, 19B Yuquan Road, Beijing 100049, China}

\author{T. P. Li}
\affil{Key Laboratory for Particle Astrophysics, Institute of High Energy Physics, Chinese Academy of Sciences, 19B Yuquan Road, Beijing 100049, China}
\affil{Department of Astronomy, Tsinghua University, Beijing 100084, China}
\affil{University of Chinese Academy of Sciences, Chinese Academy of Sciences, Beijing 100049, China}

\author{Y. Huang}
\affil{Key Laboratory for Particle Astrophysics, Institute of High Energy Physics, Chinese Academy of Sciences, 19B Yuquan Road, Beijing 100049, China}

\author{S. J. Zheng}
\affil{Key Laboratory for Particle Astrophysics, Institute of High Energy Physics, Chinese Academy of Sciences, 19B Yuquan Road, Beijing 100049, China}

\author{Z. Chang}
\affil{Key Laboratory for Particle Astrophysics, Institute of High Energy Physics, Chinese Academy of Sciences, 19B Yuquan Road, Beijing 100049, China}

\author{Y. L. Tuo}
\affil{Key Laboratory for Particle Astrophysics, Institute of High Energy Physics, Chinese Academy of Sciences, 19B Yuquan Road, Beijing 100049, China}
\affil{University of Chinese Academy of Sciences, Chinese Academy of Sciences, Beijing 100049, China}

\author{L. M. Song}
\affil{Key Laboratory for Particle Astrophysics, Institute of High Energy Physics, Chinese Academy of Sciences, 19B Yuquan Road, Beijing 100049, China}
\affil{University of Chinese Academy of Sciences, Chinese Academy of Sciences, Beijing 100049, China}

\author{Y. P. Xu}
\affil{Key Laboratory for Particle Astrophysics, Institute of High Energy Physics, Chinese Academy of Sciences, 19B Yuquan Road, Beijing 100049, China}

\author{Y. Chen}
\affil{Key Laboratory for Particle Astrophysics, Institute of High Energy Physics, Chinese Academy of Sciences, 19B Yuquan Road, Beijing 100049, China}

\author{C. Z. Liu}
\affil{Key Laboratory for Particle Astrophysics, Institute of High Energy Physics, Chinese Academy of Sciences, 19B Yuquan Road, Beijing 100049, China}

\author{Q. C. Bu}
\affil{Key Laboratory for Particle Astrophysics, Institute of High Energy Physics, Chinese Academy of Sciences, 19B Yuquan Road, Beijing 100049, China}

\author{C. Cai}
\affil{Key Laboratory for Particle Astrophysics, Institute of High Energy Physics, Chinese Academy of Sciences, 19B Yuquan Road, Beijing 100049, China}

\author{X. L. Cao}
\affil{Key Laboratory for Particle Astrophysics, Institute of High Energy Physics, Chinese Academy of Sciences, 19B Yuquan Road, Beijing 100049, China}

\author{L. Chen} 
\affil{Department of Astronomy, Beijing Normal University, Beijing 100088, China}

\author{T. X. Chen}
\affil{Key Laboratory for Particle Astrophysics, Institute of High Energy Physics, Chinese Academy of Sciences, 19B Yuquan Road, Beijing 100049, China}

\author{Y. P. Chen}
\affil{Key Laboratory for Particle Astrophysics, Institute of High Energy Physics, Chinese Academy of Sciences, 19B Yuquan Road, Beijing 100049, China}

\author{W. W. Cui}
\affil{Key Laboratory for Particle Astrophysics, Institute of High Energy Physics, Chinese Academy of Sciences, 19B Yuquan Road, Beijing 100049, China}

\author{Y. Y. Du}
\affil{Key Laboratory for Particle Astrophysics, Institute of High Energy Physics, Chinese Academy of Sciences, 19B Yuquan Road, Beijing 100049, China}

\author{G. H. Gao}
\affil{Key Laboratory for Particle Astrophysics, Institute of High Energy Physics, Chinese Academy of Sciences, 19B Yuquan Road, Beijing 100049, China}
\affil{University of Chinese Academy of Sciences, Chinese Academy of Sciences, Beijing 100049, China}

\author{Y. D. Gu}
\affil{Key Laboratory for Particle Astrophysics, Institute of High Energy Physics, Chinese Academy of Sciences, 19B Yuquan Road, Beijing 100049, China}

\author{J. Guan}
\affil{Key Laboratory for Particle Astrophysics, Institute of High Energy Physics, Chinese Academy of Sciences, 19B Yuquan Road, Beijing 100049, China}

\author{C. C. Guo}
\affil{Key Laboratory for Particle Astrophysics, Institute of High Energy Physics, Chinese Academy of Sciences, 19B Yuquan Road, Beijing 100049, China}
\affil{University of Chinese Academy of Sciences, Chinese Academy of Sciences, Beijing 100049, China}

\author{D. W. Han}
\affil{Key Laboratory for Particle Astrophysics, Institute of High Energy Physics, Chinese Academy of Sciences, 19B Yuquan Road, Beijing 100049, China}

\author{J. Huo}
\affil{Key Laboratory for Particle Astrophysics, Institute of High Energy Physics, Chinese Academy of Sciences, 19B Yuquan Road, Beijing 100049, China}

\author{S. M. Jia}
\affil{Key Laboratory for Particle Astrophysics, Institute of High Energy Physics, Chinese Academy of Sciences, 19B Yuquan Road, Beijing 100049, China}

\author{W. C. Jiang}
\affil{Key Laboratory for Particle Astrophysics, Institute of High Energy Physics, Chinese Academy of Sciences, 19B Yuquan Road, Beijing 100049, China}

\author{J. Jin}
\affil{Key Laboratory for Particle Astrophysics, Institute of High Energy Physics, Chinese Academy of Sciences, 19B Yuquan Road, Beijing 100049, China}

\author{L. D. Kong}
\affil{Key Laboratory for Particle Astrophysics, Institute of High Energy Physics, Chinese Academy of Sciences, 19B Yuquan Road, Beijing 100049, China}
\affil{University of Chinese Academy of Sciences, Chinese Academy of Sciences, Beijing 100049, China}

\author{B. Li}
\affil{Key Laboratory for Particle Astrophysics, Institute of High Energy Physics, Chinese Academy of Sciences, 19B Yuquan Road, Beijing 100049, China}

\author{C. K. Li}
\affil{Key Laboratory for Particle Astrophysics, Institute of High Energy Physics, Chinese Academy of Sciences, 19B Yuquan Road, Beijing 100049, China}

\author{G. Li}
\affil{Key Laboratory for Particle Astrophysics, Institute of High Energy Physics, Chinese Academy of Sciences, 19B Yuquan Road, Beijing 100049, China}

\author{W. Li}
\affil{Key Laboratory for Particle Astrophysics, Institute of High Energy Physics, Chinese Academy of Sciences, 19B Yuquan Road, Beijing 100049, China}

\author{X. Li}
\affil{Key Laboratory for Particle Astrophysics, Institute of High Energy Physics, Chinese Academy of Sciences, 19B Yuquan Road, Beijing 100049, China}

\author{X. B. Li}
\affil{Key Laboratory for Particle Astrophysics, Institute of High Energy Physics, Chinese Academy of Sciences, 19B Yuquan Road, Beijing 100049, China}

\author{X. F. Li}
\affil{Key Laboratory for Particle Astrophysics, Institute of High Energy Physics, Chinese Academy of Sciences, 19B Yuquan Road, Beijing 100049, China}

\author{Z. W. Li}
\affil{Key Laboratory for Particle Astrophysics, Institute of High Energy Physics, Chinese Academy of Sciences, 19B Yuquan Road, Beijing 100049, China}

\author{X. H. Liang}
\affil{Key Laboratory for Particle Astrophysics, Institute of High Energy Physics, Chinese Academy of Sciences, 19B Yuquan Road, Beijing 100049, China}

\author{J. Y. Liao}
\affil{Key Laboratory for Particle Astrophysics, Institute of High Energy Physics, Chinese Academy of Sciences, 19B Yuquan Road, Beijing 100049, China}

\author{H. W. Liu}
\affil{Key Laboratory for Particle Astrophysics, Institute of High Energy Physics, Chinese Academy of Sciences, 19B Yuquan Road, Beijing 100049, China}

\author{X. J. Liu}
\affil{Key Laboratory for Particle Astrophysics, Institute of High Energy Physics, Chinese Academy of Sciences, 19B Yuquan Road, Beijing 100049, China}

\author{X. F. Lu}
\affil{Key Laboratory for Particle Astrophysics, Institute of High Energy Physics, Chinese Academy of Sciences, 19B Yuquan Road, Beijing 100049, China}

\author{Q. Luo}
\affil{Key Laboratory for Particle Astrophysics, Institute of High Energy Physics, Chinese Academy of Sciences, 19B Yuquan Road, Beijing 100049, China}
\affil{University of Chinese Academy of Sciences, Chinese Academy of Sciences, Beijing 100049, China}

\author{T. Luo}
\affil{Key Laboratory for Particle Astrophysics, Institute of High Energy Physics, Chinese Academy of Sciences, 19B Yuquan Road, Beijing 100049, China}

\author{B. Meng}
\affil{Key Laboratory for Particle Astrophysics, Institute of High Energy Physics, Chinese Academy of Sciences, 19B Yuquan Road, Beijing 100049, China}

\author{Y. Nang}
\affil{Key Laboratory for Particle Astrophysics, Institute of High Energy Physics, Chinese Academy of Sciences, 19B Yuquan Road, Beijing 100049, China}
\affil{University of Chinese Academy of Sciences, Chinese Academy of Sciences, Beijing 100049, China}

\author{J. Y. Nie}
\affil{Key Laboratory for Particle Astrophysics, Institute of High Energy Physics, Chinese Academy of Sciences, 19B Yuquan Road, Beijing 100049, China}

\author{G. Ou}
\affil{Key Laboratory for Particle Astrophysics, Institute of High Energy Physics, Chinese Academy of Sciences, 19B Yuquan Road, Beijing 100049, China}

\author{N. Sai}
\affil{Key Laboratory for Particle Astrophysics, Institute of High Energy Physics, Chinese Academy of Sciences, 19B Yuquan Road, Beijing 100049, China}
\affil{University of Chinese Academy of Sciences, Chinese Academy of Sciences, Beijing 100049, China}

\author{R. C. Shang}
\affil{Department of Astronomy, Tsinghua University, Beijing 100084, China}

\author{X. Y. Song}
\affil{Key Laboratory for Particle Astrophysics, Institute of High Energy Physics, Chinese Academy of Sciences, 19B Yuquan Road, Beijing 100049, China}

\author{L. Sun}
\affil{Key Laboratory for Particle Astrophysics, Institute of High Energy Physics, Chinese Academy of Sciences, 19B Yuquan Road, Beijing 100049, China}

\author{Y. Tan}
\affil{Key Laboratory for Particle Astrophysics, Institute of High Energy Physics, Chinese Academy of Sciences, 19B Yuquan Road, Beijing 100049, China}

\author{W. S. Wang}
\affil{Key Laboratory for Particle Astrophysics, Institute of High Energy Physics, Chinese Academy of Sciences, 19B Yuquan Road, Beijing 100049, China}

\author{Y. D. Wang} 
\affil{Department of Astronomy, Beijing Normal University, Beijing 100088, China}

\author{Y. S. Wang}
\affil{Key Laboratory for Particle Astrophysics, Institute of High Energy Physics, Chinese Academy of Sciences, 19B Yuquan Road, Beijing 100049, China}

\author{X. Y. Wen}
\affil{Key Laboratory for Particle Astrophysics, Institute of High Energy Physics, Chinese Academy of Sciences, 19B Yuquan Road, Beijing 100049, China}

\author{B. B. Wu}
\affil{Key Laboratory for Particle Astrophysics, Institute of High Energy Physics, Chinese Academy of Sciences, 19B Yuquan Road, Beijing 100049, China}

\author{B. Y. Wu}
\affil{Key Laboratory for Particle Astrophysics, Institute of High Energy Physics, Chinese Academy of Sciences, 19B Yuquan Road, Beijing 100049, China}
\affil{University of Chinese Academy of Sciences, Chinese Academy of Sciences, Beijing 100049, China}

\author{M. Wu}
\affil{Key Laboratory for Particle Astrophysics, Institute of High Energy Physics, Chinese Academy of Sciences, 19B Yuquan Road, Beijing 100049, China}

\author{G. C. Xiao}
\affil{Key Laboratory for Particle Astrophysics, Institute of High Energy Physics, Chinese Academy of Sciences, 19B Yuquan Road, Beijing 100049, China}
\affil{University of Chinese Academy of Sciences, Chinese Academy of Sciences, Beijing 100049, China}

\author{S. Xiao}
\affil{Key Laboratory for Particle Astrophysics, Institute of High Energy Physics, Chinese Academy of Sciences, 19B Yuquan Road, Beijing 100049, China}
\affil{University of Chinese Academy of Sciences, Chinese Academy of Sciences, Beijing 100049, China}

\author{S. L. Xiong}
\affil{Key Laboratory for Particle Astrophysics, Institute of High Energy Physics, Chinese Academy of Sciences, 19B Yuquan Road, Beijing 100049, China}

\author{S. Yang}
\affil{Key Laboratory for Particle Astrophysics, Institute of High Energy Physics, Chinese Academy of Sciences, 19B Yuquan Road, Beijing 100049, China}

\author{Y. J. Yang}
\affil{Key Laboratory for Particle Astrophysics, Institute of High Energy Physics, Chinese Academy of Sciences, 19B Yuquan Road, Beijing 100049, China}

\author{Q. B. Yi}
\affil{Key Laboratory for Particle Astrophysics, Institute of High Energy Physics, Chinese Academy of Sciences, 19B Yuquan Road, Beijing 100049, China}
\affil{University of Chinese Academy of Sciences, Chinese Academy of Sciences, Beijing 100049, China}

\author{Q. Q. Yin}
\affil{Key Laboratory for Particle Astrophysics, Institute of High Energy Physics, Chinese Academy of Sciences, 19B Yuquan Road, Beijing 100049, China}

\author{Y. You}
\affil{Key Laboratory for Particle Astrophysics, Institute of High Energy Physics, Chinese Academy of Sciences, 19B Yuquan Road, Beijing 100049, China}

\author{F. Zhang}
\affil{Key Laboratory for Particle Astrophysics, Institute of High Energy Physics, Chinese Academy of Sciences, 19B Yuquan Road, Beijing 100049, China}

\author{H. M. Zhang}
\affil{Key Laboratory for Particle Astrophysics, Institute of High Energy Physics, Chinese Academy of Sciences, 19B Yuquan Road, Beijing 100049, China}

\author{J. Zhang}
\affil{Key Laboratory for Particle Astrophysics, Institute of High Energy Physics, Chinese Academy of Sciences, 19B Yuquan Road, Beijing 100049, China}

\author{W. C. Zhang}
\affil{Key Laboratory for Particle Astrophysics, Institute of High Energy Physics, Chinese Academy of Sciences, 19B Yuquan Road, Beijing 100049, China}

\author{W. Zhang}
\affil{Key Laboratory for Particle Astrophysics, Institute of High Energy Physics, Chinese Academy of Sciences, 19B Yuquan Road, Beijing 100049, China}
\affil{University of Chinese Academy of Sciences, Chinese Academy of Sciences, Beijing 100049, China}

\author{Y. F. Zhang}
\affil{Key Laboratory for Particle Astrophysics, Institute of High Energy Physics, Chinese Academy of Sciences, 19B Yuquan Road, Beijing 100049, China}

\author{H. S. Zhao}
\affil{Key Laboratory for Particle Astrophysics, Institute of High Energy Physics, Chinese Academy of Sciences, 19B Yuquan Road, Beijing 100049, China}

\author{X. F. Zhao}
\affil{Key Laboratory for Particle Astrophysics, Institute of High Energy Physics, Chinese Academy of Sciences, 19B Yuquan Road, Beijing 100049, China}
\affil{University of Chinese Academy of Sciences, Chinese Academy of Sciences, Beijing 100049, China}

\author{D. K. Zhou}
\affil{Key Laboratory for Particle Astrophysics, Institute of High Energy Physics, Chinese Academy of Sciences, 19B Yuquan Road, Beijing 100049, China}
\affil{University of Chinese Academy of Sciences, Chinese Academy of Sciences, Beijing 100049, China}


\begin{abstract}
After 34 years, the black-hole candidate EXO~1846--031 went into outburst again in 2019. We investigate its spectral properties in the hard intermediate and the soft states with {\Nu} and {\em Insight-HXMT}. A reflection component has been detected in the two spectral states but possibly originating from different illumination spectra: in the intermediate state, the illuminating source is attributed to a hard coronal component, which has been commonly observed in other X-ray binaries, whereas in the soft state the reflection is probably produced by the disk self-irradiation. Both cases support EXO~1846--031 as a low inclination system of $\sim40\degree$.
An absorption line is clearly detected at $\sim 7.2$~keV in the hard intermediate state, corresponding to a highly ionized disk wind ($\log~\xi>6.1$) with a velocity up to 0.06c. Meanwhile, quasi-simultaneous radio emissions have been detected before and after the X-rays, implying the co-existence of disk winds and jets in this system. Additionally, the observed wind in this source is potentially driven by magnetic forces.
The absorption line disappeared in the soft state and a narrow emission line appeared at $\sim6.7$~keV on top of the reflection component, which may be evidence for disk winds, but data with the higher spectral resolution are required to examine this.

\end{abstract}


\section{Introduction} \label{sec:intro}
Low-mass X-ray binaries (LMXBs) harbor a compact object, either a neutron star or a black hole, and a companion star with a mass smaller than or similar to that of the Sun (e.g. \citealt{Frank2002}). X-ray transients spend most of their lives in a quiescent state, but show bright outbursts lasting for weeks to months with a recurrence time varying from months to decades \citep{Tanaka1996,Chen1997,Tomsick2000}. Compared to super-massive black holes, stellar-mass black holes display much richer and faster variability and thus provide an ideal window to study the evolution of extreme astrophysical phenomena, such as accretion and ejection processes, in an observable timescale (e.g. \citealt{Shakura1973,Poutanen1999,Belloni2000,Uttley2005,Uttley2014,Altamirano2015,Wang2020}).

A typical outburst starts with a rapid increase of the source luminosity by several orders of magnitude, and then the luminosity decreases over a much longer time period.
Along with such an outburst, black-hole X-ray binaries (BHXBs) exhibit different characteristics in their spectra and fluxes (e.g. \citealt{Miyamoto1991,Klis1995,Mendez1997,Remillard2006}). Two main spectral states, soft and hard, have been generally distinguished in a complete cycle of an outburst. In the soft state, the spectrum shows prominent thermal emission from an accretion disk. 
In the hard state, the spectrum is dominated by non-thermal emission from a hot corona or the base of a jet (e.g. \citealt{Markoff2005}), where the soft photons from an accretion disk are inversely Compton scattered by the hot electrons of the corona. 
Additionally, an intermediate state occurs when the source transitions between these two states.

Apart from the thermal and non-thermal components, emission and absorption line features have also been widely observed in BHXBs, especially the broad emission line centered at 6.4--7~keV (e.g. \citealt{Laor1991,Miller2007,Garcia2015,Wang2018}). 
As the most prominent feature of disk reflection, such emission line is produced when the non-thermal/hard emission illuminates the accretion disk \citep{Lightman1988,George1991,Garcia2014,Dauser2014}, though the illuminating source has also been attributed to the surface emission or the boundary layer in neutron star X-ray binaries (e.g. \citealt{Ross2005,Cackett2010,Miller2015,Wang2019}). Most recently, \cite{Connors2020} found one intriguing case in which the disk reflection in the very soft state in the BHXB XTE~J1550--564 was produced by self-irradiation of the innermost part of the accretion disk.

In regard to the absorption lines discovered in BHXBs, the highly ionized ones have been attributed to the presence of disk winds, which have been suggested to be responsible for the quenching of jets (e.g. \citealt{Miller2008,Neilsen2009}). Such line features, for instance, Fe~XXV and Fe~XXVI absorption lines, are preferentially observed in systems with high inclination angles in the soft state \citep{Ponti2012}, e.g. 4U~1630--472 \citep{Diaz2014,Miller2015}, GRO~J1655$-$40 \citep{Miller2006a,Miller2008}, H~1743--322 \citep{Miller2006b,Miller2012} and GRS~1915+105 \citep{Neilsen2009,Ueda2009}. 
Theoretical studies show that such lines could be driven by radiation pressure, thermal or magnetic forces, or even the combination of these (e.g. \citealt{Proga2002,Tomaru2019}). However, the driving mechanism is still debated in accreting systems.

{\exo} was discovered on 1985 April 3, it was suggested that the source is a black-hole candidate (BHC) due to its ultrasoft spectral component observed with EXOSAT \citep{Parmar1993}. After 34 years quiescence, the {\em Monitor of All-sky X-ray Image} ({\em MAXI}/GSC) observed {\exo} being active from 2019 July 23.
A low frequency quasi periodic oscillation (QPO) at $0.71\pm0.01~$Hz in the energy band 25--250~keV has been observed with the {\em Hard X-ray Modulation Telescope} ({\hx}), dubbed {\em Insight-HXMT}, in the hard state \citep{Yang2019}. The QPO frequency increases with the source luminosity, which is a typical phenomenon observed in BHXBs. Besides that, disk reflection and disk winds have been observed in one \textit{Nuclear Spectroscopic Telescope Array} \citep[\textit{NuSTAR;}][]{Harrison2013} spectrum of this source \citep{Miller2019}. These authors report an outflowing velocity of 0.01c of the absorbing gas.

In this work, we examine the observed absorption feature with the {\Nu} observation of {\exo} in the hard intermediate state with a self-consistent photoionization model and explore its driving mechanism. We then further investigate how the outflow observed in the intermediate state, behaves in the following soft state with an {\em Insight-HXMT} observation. Finally, we discuss the origin of the absorption and emission lines in {\exo}.

\section{Observations and data analysis}
The \textit{Neil Gehrels Swift Observatory} \citep[\textit{Swift;}][]{Gehrels2004} and {\Nu} observed the outburst of EXO~1846--031 on 2019 August 2 and 3. {\em Insight-HXMT}, observed the source on 2019 September 15 (we will use {\hx}, short for {\em Insight-HXMT}, in the rest of this paper). More details about the observations are shown in Table~\ref{tab:obs}.

The {\em Swift}/XRT data were taken in the Windowed Timing mode. We reduced the data using {\sc xrtpipeline} following the official user's guide \footnote{\url{https://swift.gsfc.nasa.gov/analysis/xrt\_swguide\_v1\_2.pdf}}.
Using the {\Nu} Data Analysis Software (NuSTARDAS v2.0.0), we created the {\Nu} spectra and lightcurves of the source and the background with two circular extraction regions of 80\arcsec, one centered at the source and another one centered away from the source. The {\Nu} CALDB (v20200912) is applied in this work.

We extracted the {\hx} lightcurves and spectra by following the official user guide \footnote{\url{http://enghxmt.ihep.ac.cn/sjfxwd/169.jhtml}} and using the latest software {\sc HXMTDAS} (v2.02) \citep{Zhang2020}. The criteria for screening good time intervals is as follows: 1) the elevation angle $>$10\degree; 2) the geomagnetic cutoff rigidity $>$8~GeV; 3) the pointing offset angle $<$ 0.1\degree; 4) at least 300~s away from the South Atlantic Anomaly (SAA). The background was estimated using the scripts {\sc LEBKGMAP} and {\sc MEBKGMAP} \citep{Liao2020, Guo2020}. The energy bands used in the spectral analysis are 1-9\,keV and 10-20\,keV for the low and medium energy (LE and ME) detectors, respectively \citep{Chen2020, Cao2020}.

The spectra were grouped into bins to have at least 30 counts \citep{Cash1979}.
To account for the neutral hydrogen column absorption of the interstellar medium, we employed the \texttt{tbabs} model and the {\sc wilm} solar abundance \citep{Wilms2000} in xspec (v12.11.1). In this paper, all uncertainties are quoted at 1-$\sigma$ confidence level.

\begin{table} 
\caption{{\sw}, {\Nu} and {\hx} observations of {\exo} used in this paper.}
\renewcommand{\arraystretch}{1.3}
\footnotesize
\centering
\begin{tabular}{lcccccc}
\hline \hline
Mission/Instrument & ObsID & MJD & Exposure (ks)\\
{\sw}/XRT&00011500002&58697.3& 1.1 \\
\hline
{\Nu}/FPMA&90501334002&58698.1& 22.2\\
{\Nu}/FPMB& && 22.6\\
\hline
{\hx}/LE&P0214050022&58741.1&13.8\\
{\hx}/ME&&&24.6\\
\hline
\end{tabular}\label{tab:obs}

\end{table}

\begin{figure}
    \centering
    \includegraphics[width=1\linewidth]{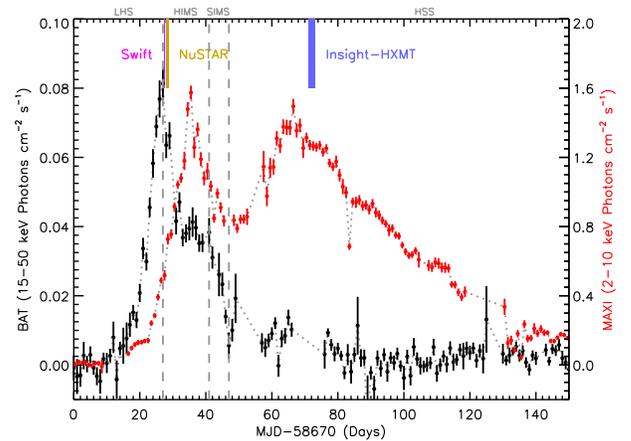}
    \caption{Long-term lightcurves of EXO~1846--031 during its 2019 outburst observed with {\em Swift}/BAT (15--50~keV, black points) and {\em MAXI} (2--10~keV, red points). The magenta, yellow and blue bars represent the time of the {\sw}, {\Nu} and {\hx} observations. The gray dashed lines indicate the spectral states of EXO~1846--031, taken from \protect{\cite{Liu2020}}. `LHS', `HIMS', `SIMS' and `HSS' are short for low hard, hard intermediate, soft intermediate and high soft states, respectively.}
    \label{fig:lc}
\end{figure}

\section{analysis and results}
Fig.~\ref{fig:lc} shows the long-term light curves in the 2--10\,keV and 15--50\,keV bands observed with \textit{MAXI} and \textit{Swift}/BAT in red and black, respectively.
It is clear that the peak of hard X-rays precedes that of soft X-rays, showing the canonical evolution of outbursts in LMXBs \citep[e.g.][]{Remillard2006,Belloni2010,MunozDarias2011}.
\cite{Liu2020} conducted a timing analysis of {\exo} in the same outburst with {\hx} and {\em Neutron star Interior Composition Explorer} ({\em NICER}). Based on the relative changes in the hardness ratio diagram and the fractional rms integrated in the $2^{-5}-32$~Hz, they classify the entire outburst into four spectral states, i.e. low hard, hard intermediate, soft intermediate and high soft (see the gray dashed lines in Fig.~\ref{fig:lc}). According to their classification, our \textit{Swift}/XRT and {\Nu} observations took place in the hard intermediate state and the {\hx} observation took place in the soft state. Although it is unclear when the source left the soft state in the end because of the low count rate, this does not affect the results present in this paper.

\begin{figure}
    \centering
    \includegraphics[width=1\linewidth]{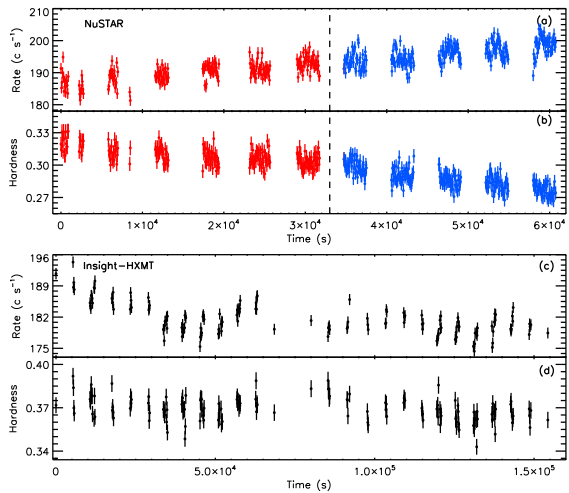}
    \caption{Upper panels: the count rate of both modules in the 3--79~keV band and the hardness ratio (10--79~keV/3--10~keV) during the {\Nu} observation of {\exo}. The dashed line separates the observation into low-flux (red) and high-flux (blue) segments. Lower panels: the count rate in the 1--10~keV band and the hardness ratio (4--10~keV/1--4~keV) during the {\hx} observation.}
    \label{fig:hardness}
\end{figure}

We extracted the {\Nu} background-subtracted light curves in the  3--10\,keV and 10--79\,keV energy bands (Fig.~\ref{fig:hardness}) and calculated the hardness ratio.
It is clearly shown that the count rate (the summation of FPMA and FPMB) gradually increases with time by around 10\%, accompanied by a reduction of the hardness, i.e. a spectral softening. We hence separated the {\em NuSTAR} spectrum into two segments with similar counts, low-flux and high-flux, as the dashed line in Fig.~\ref{fig:hardness} indicates. As to the background-subtracted {\hx} light curve, the hardness ratio between the light curves in the 4--10~keV and 1--4~keV bands shows no clear trend with time (see the bottom panel of Fig.~\ref{fig:hardness}). We thus conduct a time-average spectral analysis with the {\hx} observation.

\subsection{Spectral analysis}
Since the {\sw}, {\Nu} and {\hx} observations caught {\exo} in different spectral states, we analyzed them separately and applied different models to fit the spectra with individual requirements.

\subsubsection{The hard component dominated state} \label{sec:nu}
We fitted the {\em Swift}/XRT spectrum with an absorbed power law, \texttt{tbabs*powerlaw}, in the energy band 1--10~keV and obtained a column density, $N_{\rm H}=6\pm0.1\times10^{22}$~cm$^{-3}$, a photon index, $\Gamma=1.55\pm0.03$ and the normalization of the component \texttt{powerlaw}, $1.0\pm0.1$, with $\chi^2=615.5$ for 588 dof.  
Adding a soft component, \texttt{diskbb}, to the model improves the fit by $\Delta \chi^2=5.32$ for 2 dof fewer whereas the F-test probability of 0.08 indicates that this component is not significantly required by the data. 

\begin{figure}
    \centering
    \hspace{-0.5cm}
    \mbox{\includegraphics[angle=270,width=1\linewidth]{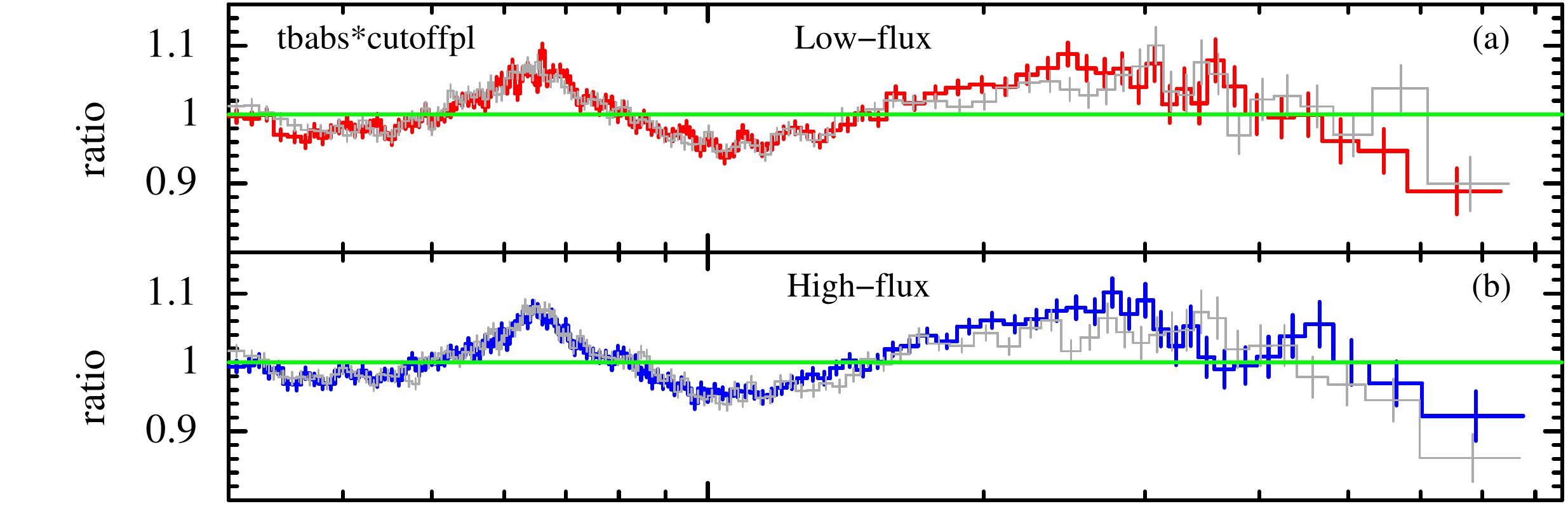}}
    \vspace{-0.05cm}
        \hspace{-0.5cm}
    \mbox{\includegraphics[angle=270,width=1\linewidth]{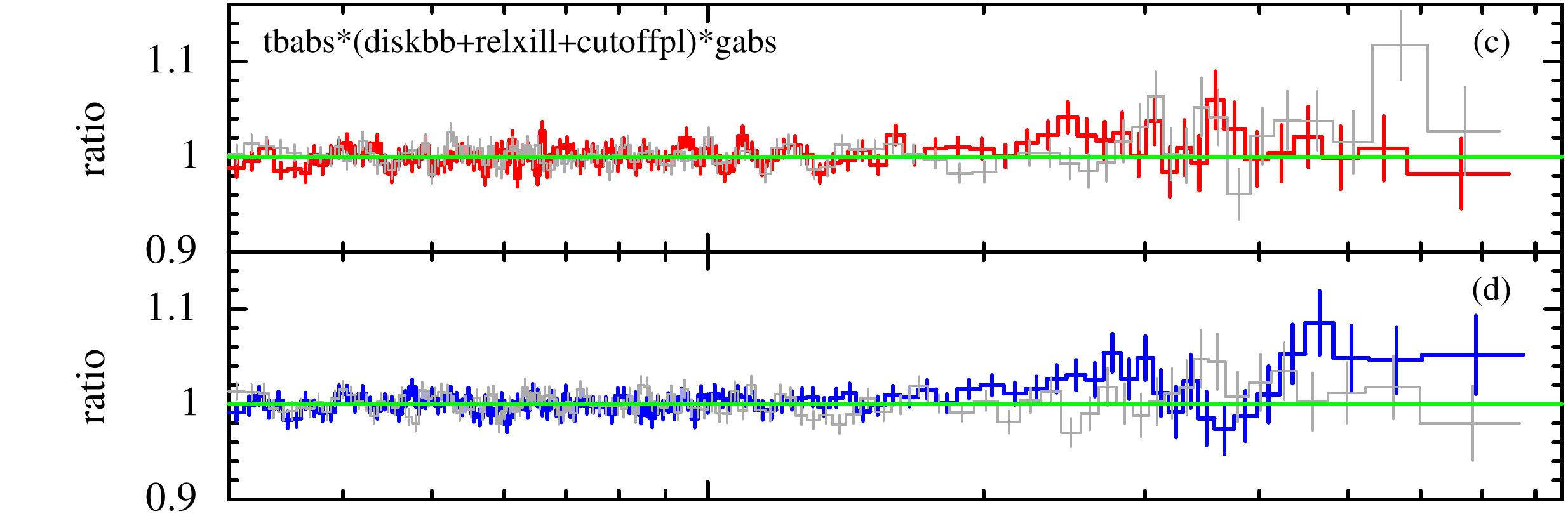}}
     \vspace{-0.05cm}
         \hspace{-0.5cm}
    \mbox{\includegraphics[angle=270,width=1\linewidth]{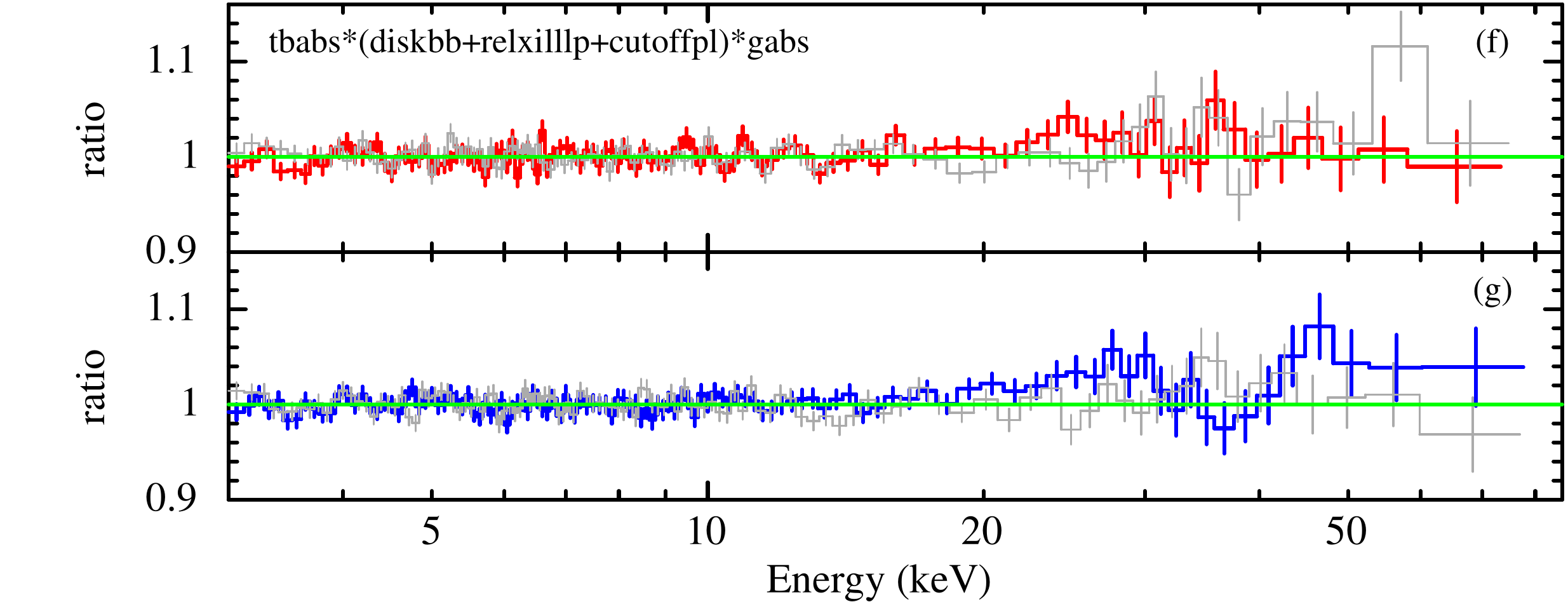}}
    
    \caption{Ratio residuals of the 3--78~keV {\Nu} spectra of {\exo} to different models. The red and blue lines represent the FPMA spectra in the low- and high-flux, respectively. The FPMB spectra are shown in gray for reference.}
    \label{fig:nu_pha}
\end{figure}

We jointly fitted the 3--79~keV FPMA and B continuum of the two segments of the {\Nu} observation with an absorbed power law with a high-energy cutoff \texttt{cutoffpl}. 
We fixed the value of the column density at the one derived from the fit to the XRT spectrum, $N_{\rm H}=6\times10^{22}$~cm$^{-3}$, and obtained a fit with $\chi^2=8746.8$ for 4149 dof. 
Figs.~\ref{fig:nu_pha}a and b show the ratio residuals of the spectra to the continuum model \texttt{tbabs*cutoffpl}. The residuals show a broad emission line centered at 6--7~keV and a hump peaking at 20--30~keV, indicating the presence of a reflection spectrum.
We thus continued our spectral analysis by including physical reflection models \texttt{relxill(lp)} (v1.3.9\footnote{\url{http://www.sternwarte.uni-erlangen.de/~dauser/research/relxill/}}) in which the reflection component is produced by the irradiation of the hard component.

In both models, \texttt{relxill} and \texttt{relxilllp}, we adopted the outer radius, $R_{\rm out}=400~R_{\rm g}$ ($R_{\rm g}$ = $GM/c^2$). We assumed that the spin parameter, $a_{*}$, the inclination of the system, $i$, and the iron abundance, $A\rm_{Fe}$, of the accretion disk do not vary on the time scales of the observation and hence we linked these parameters during the fit to be the same in the two segments. We linked the outer emissivity index to the inner one, $q_{\rm out}=q_{\rm in}$, to vary together in \texttt{relxill}. Compared to \texttt{relxill}, \texttt{relxilllp} assumes a lamppost geometry for the illuminating source. When we fitted this component, the spin parameter could not be constrained by the data. We thus fixed it at the upper limit, i.e. $a_{*}=0.998$, allowing the maximum space parameter for the disk inner radius.

We got a reasonably good fit by including the reflection component, e.g. in the fit using \texttt{relxill}, the best-fit is $\chi^2=4329.2$ for 4132 dof. 
Meanwhile there is a soft excess present at below 5~keV. Adding the component \texttt{diskbb} to the model improves the fit with $\Delta \chi^2=103.8$ for 4 dof fewer. We therefore include this component in our model.

\begin{figure}
    \centering
    \mbox{\includegraphics[angle=270,width=1\linewidth]{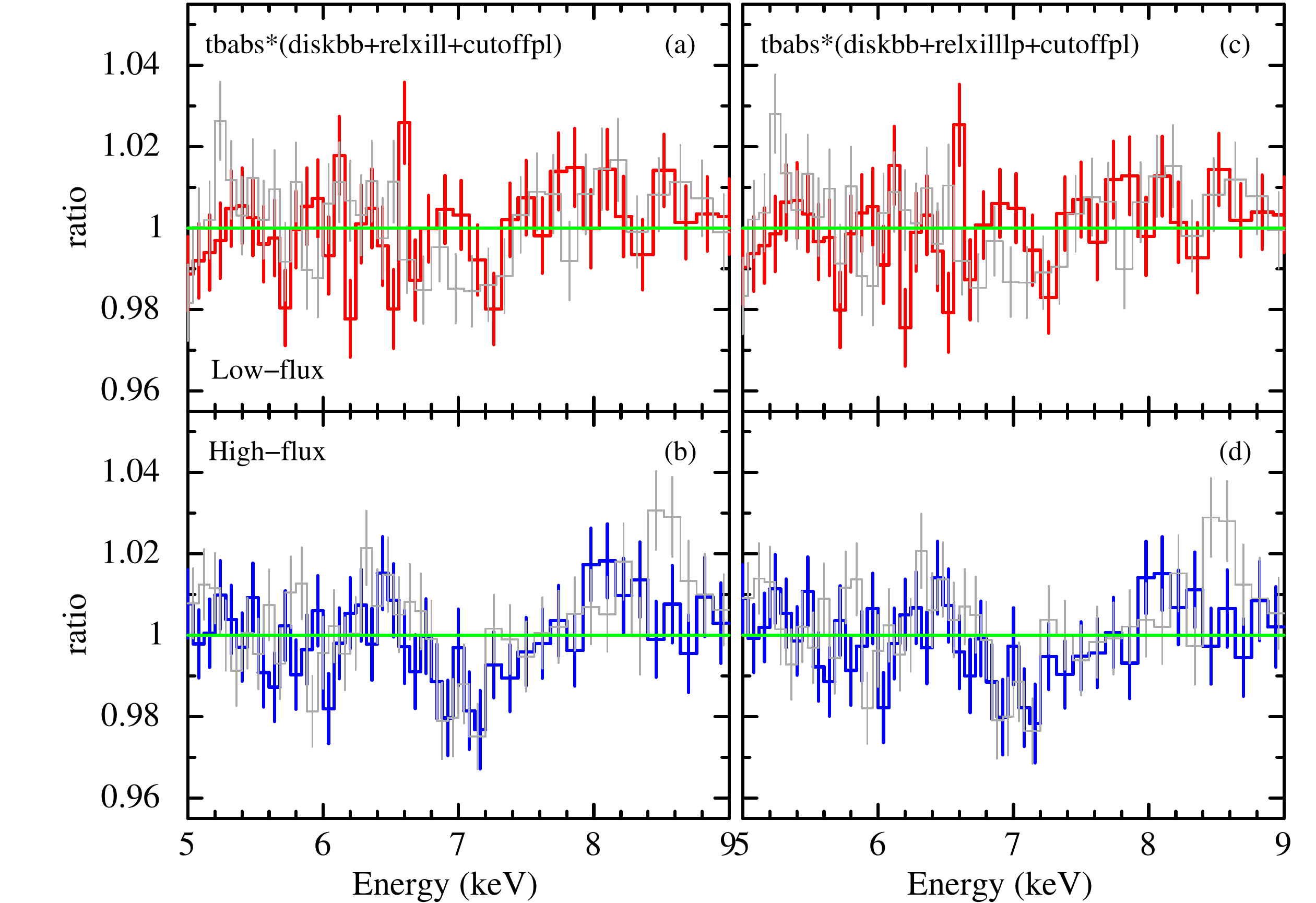}}
    \hspace{1cm}
    \caption{Ratio residuals of the 5--9~keV {\Nu} spectra of {\exo} to different models. The colors are defined the same as in Fig.~\ref{fig:nu_pha}.}
    \label{fig:nu_gabs}
\end{figure}

Additionally, a moderately broad absorption feature appears in the spectra (see Fig.~\ref{fig:nu_gabs}). In both segments, the absorption feature is independent of the choice of the reflection model. For the high-flux segment, the strength of the absorption feature is consistent in both the FPMA and B spectra, whereas for the low-flux segment, the feature in the FPMA spectrum is subtle. To test if an absorption line is required in the fit of the low-flux segment, we added an absorption line component, \texttt{gabs}, at $\sim 7.2$~keV to the model. The component \texttt{gabs} cannot be well constrained if we let its parameters vary between the FPMA and B spectra. We hence linked the parameters of \texttt{gabs} between the two spectra. The fit to the low-flux spectra improved by $\Delta \chi^2=32.1$ for 3 dof fewer. We thus included \texttt{gabs} in our model and the overall fit to the four spectra significantly improved, for instance, $\Delta \chi^2=103.3$ for 6 dof fewer in the fit with \texttt{relxill}. We obtained the best-fitting reflection fraction, for the low- and high-flux segments, of $0.10\pm0.01$ and $0.11\pm0.01$ in \texttt{relxill} and $0.26\pm0.04$ and $0.29\pm0.02$ in \texttt{relxilllp}, respectively. In order to calculate the individual flux of the reflection and the power law, we set the reflection fraction to a negative value which makes \texttt{relxill} and \texttt{relxilllp} only account for the reflection component.

We show the best-fitting parameters and the individual fluxes, $F\rm_{dbb}$, $F\rm_{rel}$, $F\rm_{pl}$, and total flux, $F\rm_{tt}$, in Table~\ref{tab:nu}. Ratio plots of the models are shown in Fig.~\ref{fig:nu_pha}. The fits of the model with either \texttt{relxill} or \texttt{relxilllp} are equally good and the common parameters in both models are consistent. The inclination angle of the accretion disk ranges from $37\degree$ to $44\degree$ and the iron abundance is about $5$ in the units of solar abundance.
The inner radius of the accretion disk is weakly constrained, $R_{\rm in}=1-11~R_{\rm g}$ in the low-flux segment and $R_{\rm in}=1-84~R_{\rm g}$ in the high-flux segment. The photon index slightly increases as the total flux increases. Additionally, the large coronal height provided by the component \texttt{relxilllp} probably explains why the spin of this source is hard to be constrained \citep{Dauser2013,Fabian2014}.

\begin{table*} 
\caption{Best-fitting parameters of the two {\Nu} segments of {\exo} with the models \texttt{tbabs*(diskbb+relxill+cutoffpl)*gabs} and  \texttt{tbabs*(diskbb+relxilllp+cutoffpl)*gabs} in the energy band 3--79~keV.}
\renewcommand{\arraystretch}{1.2}
\setlength{\tabcolsep}{2pt}
\footnotesize
\centering
\begin{tabular}{clcccccc}
\hline \hline
\multicolumn{2}{c}{\multirow{2}{*}{Components }}& \multicolumn{2}{c}{\texttt{relxill}} & \multicolumn{2}{c}{\texttt{relxilllp}} \\
& & low-flux  &  high-flux  & low-flux  &  high-flux  \\
\texttt{tbabs} &$N\rm_{H}~(10^{22}~cm^{-2})$ &\multicolumn{2}{c}{$6.0^{f}$} &\multicolumn{2}{c}{$6.0^{f}$}\\
\hline
\texttt{diskbb}&$T\rm_{dbb}$~(keV)& $0.40\pm0.04$ &$0.52\pm0.05$ & $0.41\pm0.04$ & $0.52\pm0.06$\\
&$N\rm_{dbb}$~$(R_{\rm dbb}^2/D_{10}^2 \cos~i)$&$11849.7_{-6830.4}^{+18985}$ &$1772.2_{-819.8}^{+1961.5}$&$9816.9_{-5478.3}^{+17516.2}$&$1582.6_{-779.1}^{+1522.9}$  \\
\noalign{\smallskip}
\hline                              
\texttt{relxill(lp)}& $q\rm_{in}$ &$2.6\pm0.4$&$1.9\pm0.3$ & - & -\\
& $h~(R\rm_{g})$ &- & - &$10.2\pm5.2$&$21.0_{-19.0}^{+478.4}$ \\
& $A\rm_{Fe}~(A_{\rm Fe_{\odot}})$ &\multicolumn{2}{c}{$5.0\pm0.6^{l}$}&\multicolumn{2}{c}{$5.0\pm0.5^{l}$}\\
& $i$~($^{\circ}$) &\multicolumn{2}{c}{$41.4_{-1.5}^{+2.6}$}&\multicolumn{2}{c}{$38.8_{-2.2}^{+1.6l}$}\\
& $a_{*}$ &\multicolumn{2}{c}{$0.998^{f}$}&\multicolumn{2}{c}{$0.998^{f}$}\\
&$\Gamma$&$1.75\pm0.01$ &$1.81\pm0.01$ &$1.76\pm0.01$ &$1.82\pm0.01$ \\
&$E_{\rm cut}$~(keV) &$62.5_{-2.0}^{+2.5}$ &$63.2\pm2.6$&$62.0_{-3.5}^{+2.4}$ &$65.8_{-1.3}^{+2.5}$  \\
&Refl\_frac&$-1^f$ &$-1^f$ &$-1^f$ &$-1^f$  \\
&$R\rm_{in}$~($R\rm_{g})$&$3.1_{1p}^{+2.7}$&$1.8_{1p}^{+19.0}$&$3.3_{1p}^{+7.8}$ &$59.5_{-17.9}^{+24.8}$ \\
&$\log \xi$~($\rm erg~cm~s^{-1}$)&$3.71_{-0.12}^{+0.04}$ &$3.82\pm0.05$ &$3.72_{-0.11}^{+0.03}$ &$3.91\pm0.05$ \\
&$N\rm_{rel(lp)}~(10^{-3})$&$2.0\pm0.2$ &$2.2\pm0.2$&$8.2\pm1.5$ &$6.7_{-1.1}^{+3.4}$\\ 
\noalign{\smallskip}
\hline                              
\texttt{cutoffpl}& $N_{\rm pl}$ &$1.2\pm0.1$&$1.3\pm0.1$ & $1.2\pm0.1$ & $1.3\pm0.1$\\
\hline
\texttt{gabs}&$E_{\rm gabs}$~(keV)&$7.25\pm0.05$&$7.24\pm0.05$&$7.24\pm0.04$&$7.23\pm0.05$\\
&$\sigma\rm_{gabs}$~(keV)&$0.17\pm0.10$&$0.31\pm0.04$ & $0.17\pm0.09$&$0.37\pm0.07$\\
&strength ($10^{-2}$)&$2.0\pm0.6$ &$3.6\pm0.6$ &$2.0\pm0.4$ &$4.1\pm1.0$ \\
\hline
\texttt{const}&FPMA & \multicolumn{2}{c}{$1^f$} &\multicolumn{2}{c}{$1^f$}\\
&FPMB&\multicolumn{2}{c}{$1.003\pm0.001$}&\multicolumn{2}{c}{$1.003\pm0.001$}\\
\hline
\multicolumn{2}{c}{$\chi^{2}$/d.o.f.} &\multicolumn{2}{c}{4122.2/4122}&\multicolumn{2}{c}{4124.7/4122}\\
\hline
&$F\rm_{dbb}$&$0.4\pm0.1$ &$0.4\pm0.1$ &$0.4\pm0.1$ &$0.4\pm0.1$\\
&$F\rm_{rel}$&$1.2\pm0.1$ &$1.6\pm0.2$&$1.2\pm0.1$ &$1.6\pm0.1$ \\
&$F\rm_{pl}$&$4.2\pm0.1$ &$4.2\pm0.2$&$4.2\pm0.1$ &$4.2\pm0.1$ \\
&$F\rm_{tt}$&$5.9\pm0.1$ &$6.2\pm0.1$&$5.8\pm0.1$ &$6.2\pm0.1$ \\
\hline
\end{tabular}
\begin{flushleft}
{\bf Note:} The first and the second groups are corresponding to the fits with \texttt{relxill} and \texttt{relxilllp}, respectively. The symbol $l$ indicates that the parameters are linked to vary across the observations; $p$ means that the parameter pegs at its limit; $f$ means that the parameter is fixed during the fit. All the unabsorbed fluxes, $F\rm_{dbb}$, $F\rm_{rel}$, $F\rm_{pl}$ and $F\rm_{tt}$, are calculated in the energy band of 2$-$10~keV in units of $10^{-9}~\rm erg~cm^{-2}s^{-1}$. 
\end{flushleft}    \label{tab:nu}
\end{table*}

\begin{figure}
    \centering
    \mbox{\includegraphics[angle=270,width=0.9\linewidth]{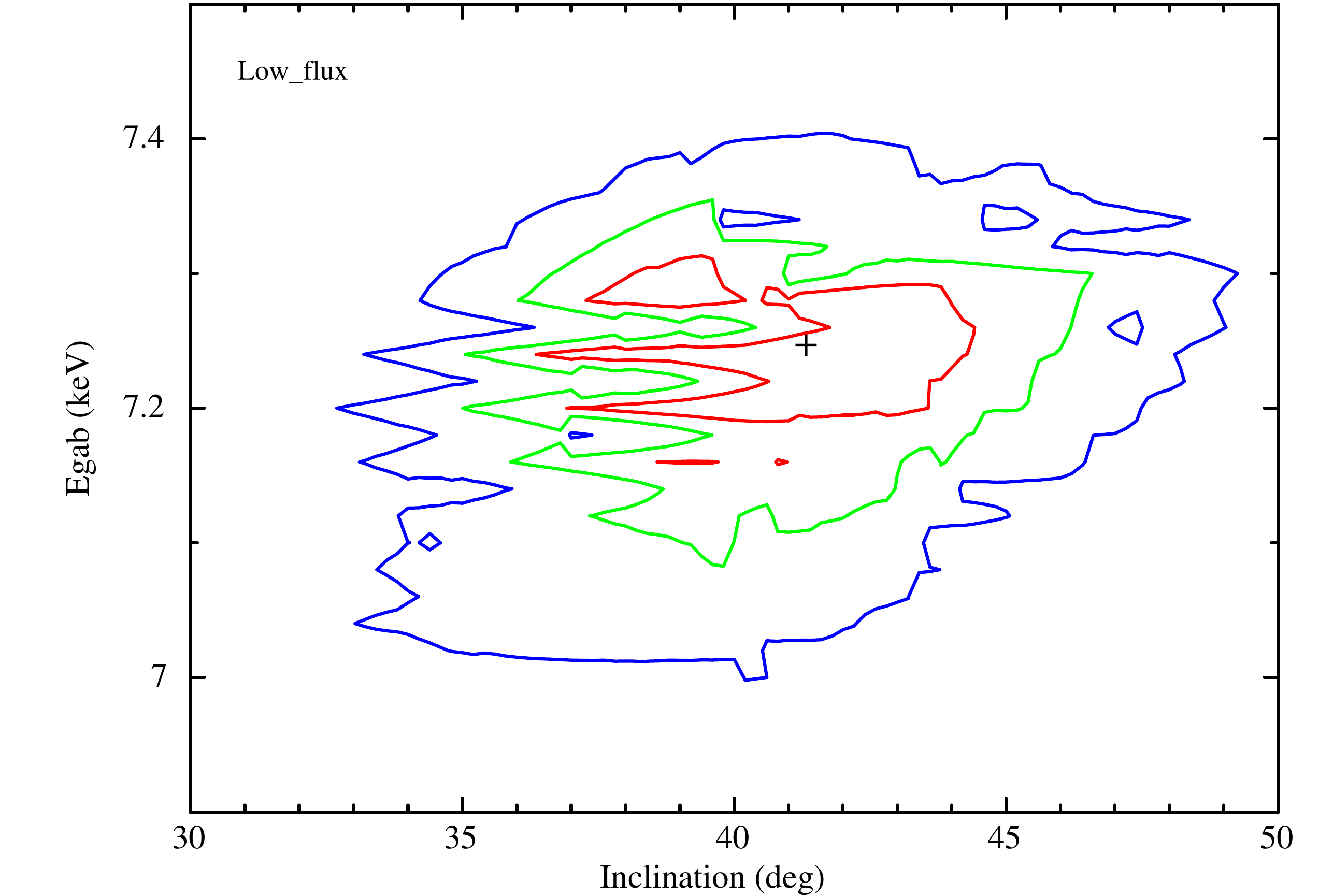}}
    \vspace{0.1cm}
    \mbox{\includegraphics[angle=270,width=0.9\linewidth]{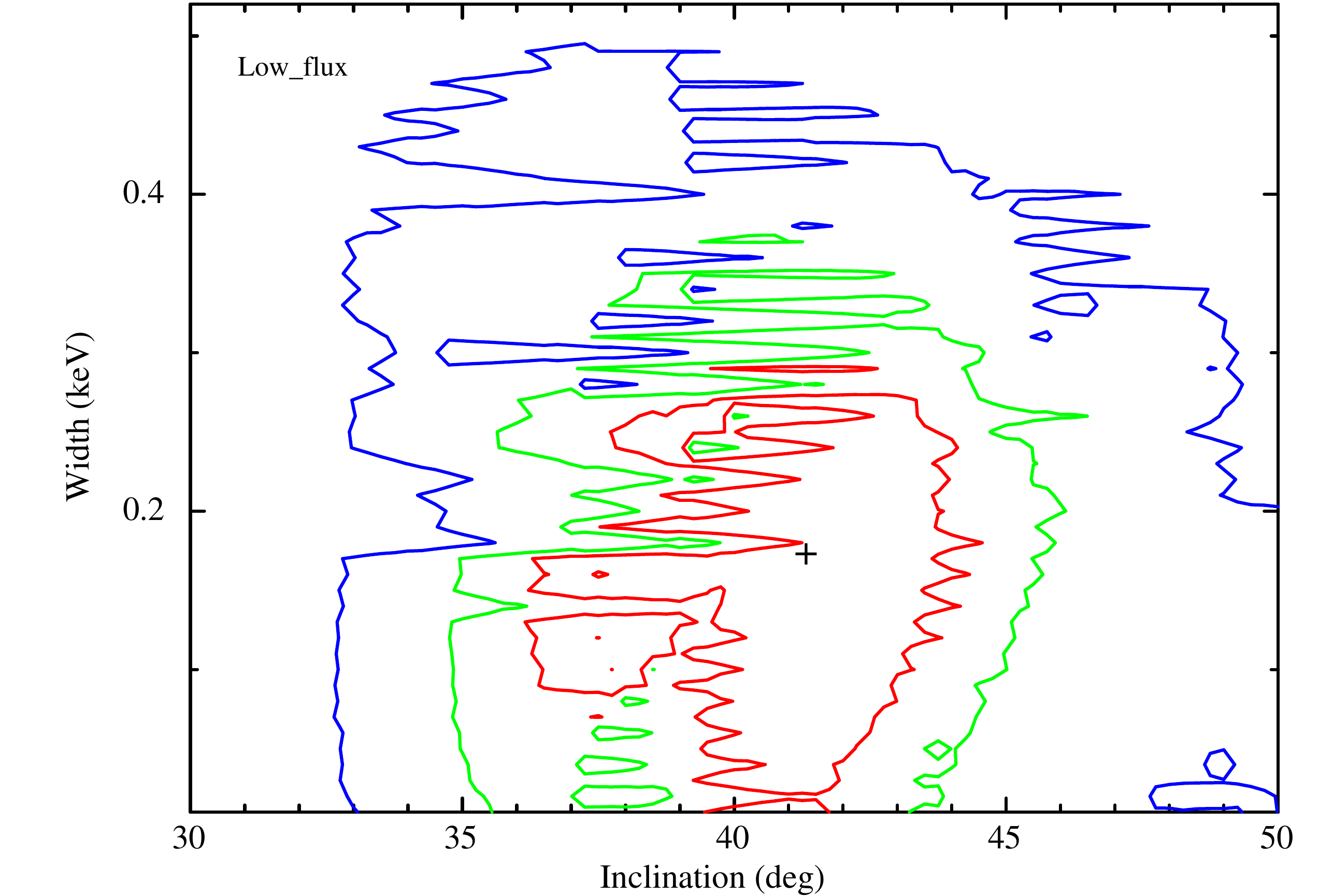}}
    \vspace{0.1cm}
    \mbox{\includegraphics[angle=270,width=0.9\linewidth]{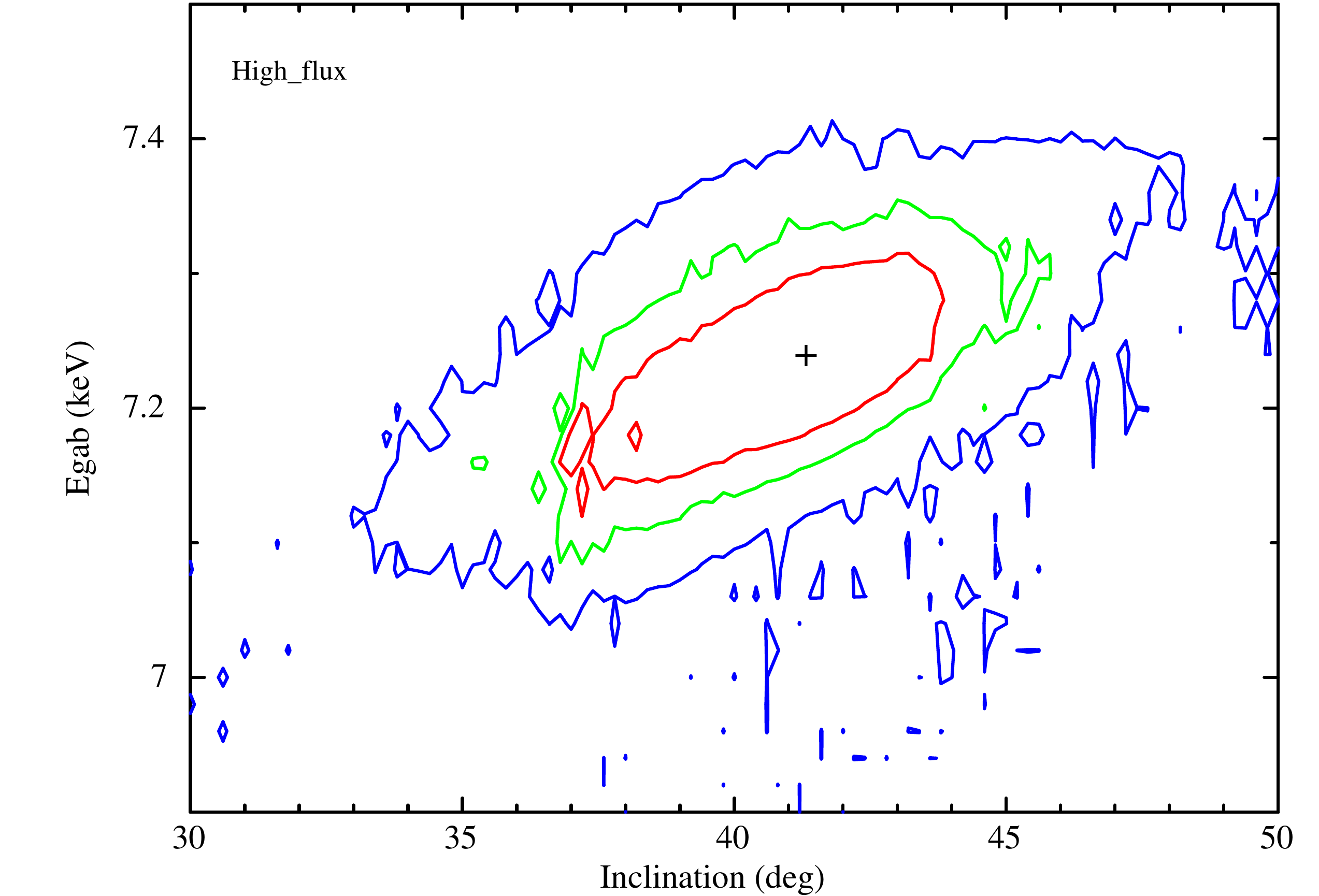}}
    \vspace{0.1cm}
    \mbox{\includegraphics[angle=270,width=0.9\linewidth]{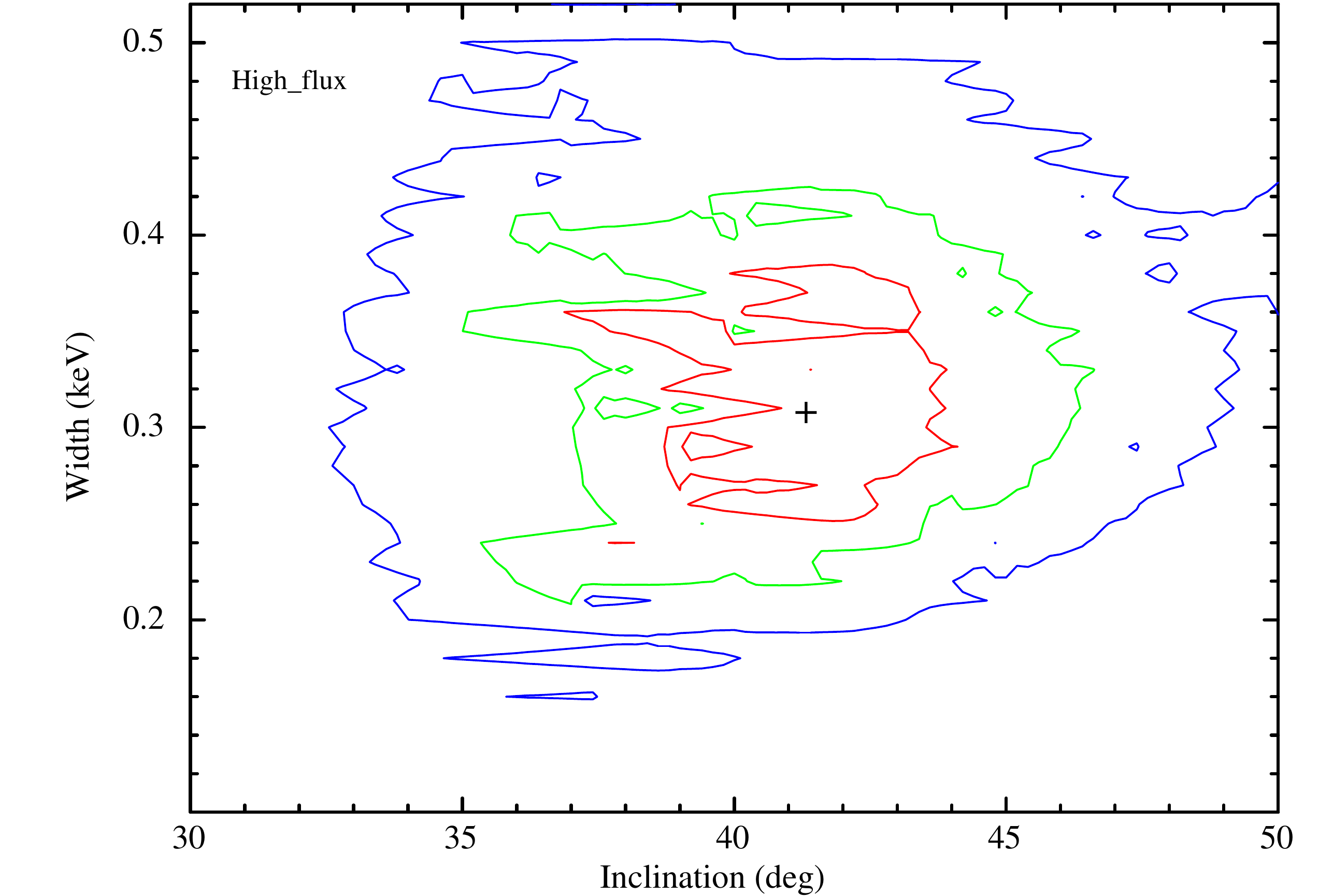}}
    \caption{Contour plots for the energy or the width of the absorption line vs disk inclination at the 68\% (red), 90\% (green) and 99\% (blue) confidence levels. The best-fitting values of these parameters are marked with a cross.}
    \label{fig:nu_contr}
\end{figure}

\cite{Miller2019} have reported the presence of an absorption line in {\exo} as well as a higher inclination angle, $75\degree$, than ours. To check if we have obtained the inclination angle at a clear global minimum and, to assess a possible correction between the absorption line and the inclination, we plotted $\chi^2$ confidence contours between the inclination and the line energy or width of the \texttt{gabs} component. Fig.~\ref{fig:nu_contr} shows that although the disk inclination is correlated to the energy of the \texttt{gabs}, the inclination angle is independent of the width of the \texttt{gabs} and is lower than $45\degree$ at $90\%$ confidence level.
If we freeze the inclination angle at $75\degree$, the fit worsens with $\Delta \chi^2=26.5$ for 1 dof fewer. We provide a further comparison at the end of Section~\ref{sec:disk_wind}.

To explore the physical properties of the absorbing gas which produces the absorption features, we created a grid of photoionization models with {\sc XSTAR} \citep{Kallman2001} using a power-law spectral input form with $\Gamma=1.8$ based on the fit to the {\Nu} observation.
For the other parameters of this grid of models, we set fixed
values of a gas density at $n_{\rm gas} = 10^{15}~\rm cm^{-3}$ and a covering fraction at $f=\Omega/4\pi= 0.5$, following the assumptions for disk winds in LMXBs \citep{Miller2015}, a source luminosity at $L =2.2 \times 10^{38}~\rm erg~s^{-1}$ in the energy band 0.01--100~keV (a distance of 8~kpc is adopted from  \citealt{MillerJones2019}). We calculated the turbulent velocity of the outflow with the best-fitting energy and width of \texttt{gabs} and obtained $v_{\rm turb}=7000\pm4100~\rm km~s^{-1}$ for the low-flux segment and $v_{\rm turb}=13000\pm1700~\rm km~s^{-1}$ for the high-flux segment. 
We hence adopted an average turbulent velocity of $10000~\rm km~s^{-1}$ as the input. 
The variable parameters are the iron abundance, $A_{\rm Fe}$, the equivalent hydrogen column density, $N_{\rm gas}$, ionization parameter, $\xi_{\rm gas}$ and the shift velocity, $v$. The grid was read into {\sc xspec} as a multiplicative model to fit the {\Nu} spectra, replacing the line model \texttt{gabs}, in which the iron abundance in the grid photoionization model was linked to that in the reflection model.  

The absorption feature is well fit with the precalculated grid. We show the best-fitting parameters of the grid in Table~\ref{tab:xstar}. The rest of the parameters obtained here are consistent with those obtained from the fit with \texttt{gabs}. The measured parameters of the low-flux segment are not constrained as well as those of the high-flux, but both suggest a highly ionized and fast outflowing wind in {\exo}. 

Once the ionization parameter is obtained, we can use it to calculate the launching radius of the disk winds and the mass outflow rate.
We define the launching radius of disk winds as $R_{\rm launch}=\sqrt{L/n_{\rm gas}\xi_{\rm gas}}$ where $L$ is the ionizing luminosity, $n_{\rm gas}$ is the gas density and $\xi_{\rm gas}$ is the ionization parameter \citep{1969Tarter,Kallman2001}. Regarding the wind mass outflow rate, we estimated it through $\dot{M}_{\rm wind}=\Omega \mu m_{\rm p} v L/\xi_{\rm gas}$, where $\Omega$ is the opening angle ($\Omega=2\pi$ is assumed), $\mu$ is the average molecular weight of the gas ($\mu=1.23$ is assumed as in some other black-hole systems, e.g. \citealt{Janiuk2015,Trueba2019}), $m_{\rm p}$ is the mass of a proton and $v$ is the outflow velocity. The mass accretion rate is roughly estimated as $\dot{M}_{\rm acc}=L/(\eta c^2)$ where the efficiency $\eta=0.1$ is assumed.
The calculated launching radius and the ratio of the mass outflow rate over the mass accretion rate are also listed in Table~\ref{tab:xstar}.
Overall, all the parameters of the photoionization model of the two segments are consistent within errors.

\begin{table}
\caption{The assumed input and measured output parameters of the {\sc xstar} model applied to the {\Nu} segments and the derived wind properties in {\exo}.}
\renewcommand{\arraystretch}{1.3}
\setlength{\tabcolsep}{2pt}
\footnotesize
\centering
\begin{tabular}{llccc}
\hline \hline
\multicolumn{2}{c}{\texttt{xstar} } & low-flux &  high-flux \\
\hline
input & $\Gamma$&\multicolumn{2}{c}{-1.8}\\
&$L~(10^{38}~ \rm erg~s^{-1})$&\multicolumn{2}{c}{2.2}\\
 &$\log n_{\rm gas}~(\rm cm^{-3})$&\multicolumn{2}{c}{15}\\
&$f=\Omega/4\pi$&\multicolumn{2}{c}{0.5}\\
&$v_{\rm turb}~(\rm km~s^{-1})$&\multicolumn{2}{c}{10000}\\
\hline
output&$N_{\rm gas}~(\rm 10^{22}~\rm cm^{-2})$&$1.61_{-0.17}^{+1.79}$&$1.71_{-0.15}^{+0.09}$\\
&$\log \xi_{\rm gas}~(\rm erg~cm~s^{-1})$& $6.40_{-0.15}^{+0.96}$ &$6.23\pm0.04$\\
&$v/c$&$-0.05\pm0.02$&$-0.06\pm0.01$\\
\hline
measurement&$R_{\rm launch}~(\rm 10^{10}~cm)$&$0.03\pm0.02$&$0.04\pm0.002$\\
&$\dot{M}\rm_{wind}~(10^{18}~g~s^{-1})$&$1.70_{-0.59}^{+3.75}$&$3.01\pm0.28$\\
&$\dot{M}_{\rm wind}/\dot{M}_{\rm acc}$&$0.69_{-0.24}^{+1.53}$&$1.23\pm0.11$\\
\noalign{\smallskip}
\hline
\end{tabular}
\begin{flushleft}
\end{flushleft}    \label{tab:xstar}
\end{table}

\subsubsection{The soft component dominated state}\label{sec:hx}
\begin{figure}
    \centering
    \hspace{-0.5cm}
    \includegraphics[angle=270,width=1\linewidth]{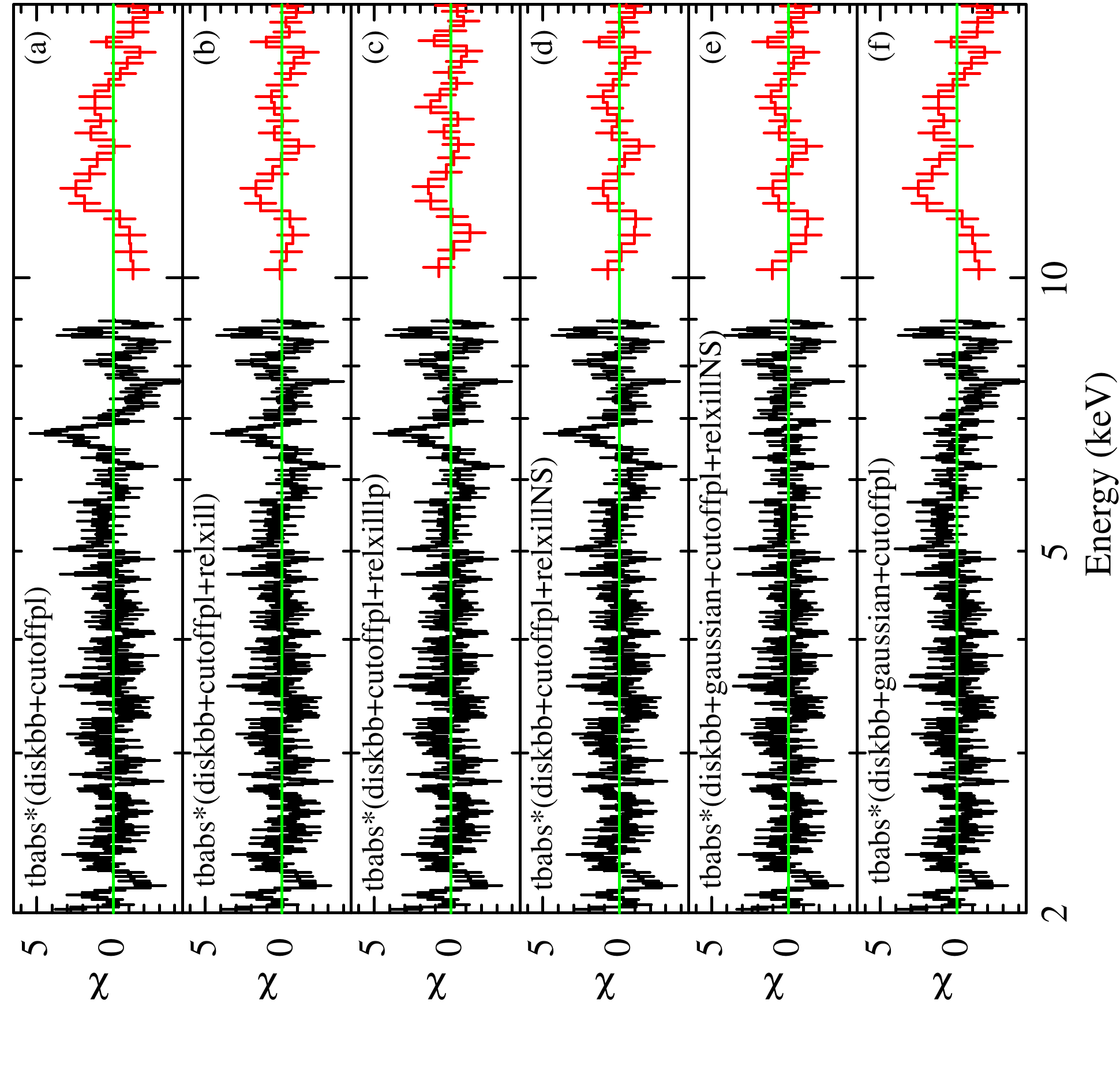}
    \caption{Residuals of the {\hx} spectra to different models. The black and red lines represent the LE and ME spectrum, respectively.}
    \label{fig:ra_hxmt}
\end{figure}

For the {\hx} observation, we jointly fitted the continuum of the LE and ME spectra in the energy bands 2--9~keV and 10--20~keV with the model \texttt{tbabs*(diskbb+cutoffpl)}. This gives a fit with $\chi^2=1053.6$ for 991 dof.
Surprisingly, the spectrum in the soft state does not show the need of an absorption line as the source in the intermediate state, but requires a narrow emission line around 6.7~keV (see the residuals in Fig.~\ref{fig:ra_hxmt}a). Besides that, there seems to be a weak excess appearing at above 10~keV. 

\begin{figure}
    \centering
    \mbox{\includegraphics[angle=270,width=0.9\linewidth]{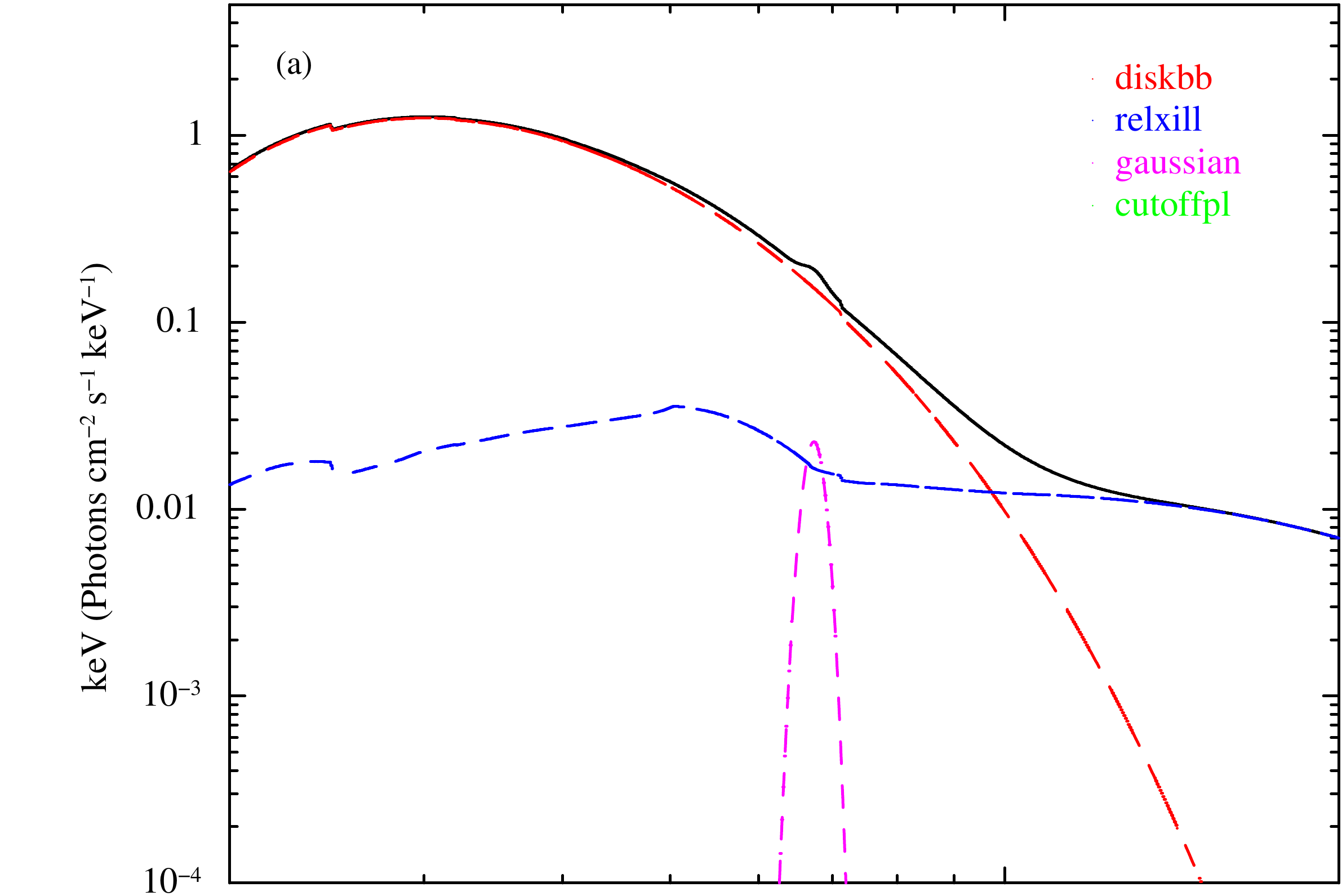}}
    \mbox{\includegraphics[angle=270,width=0.9\linewidth]{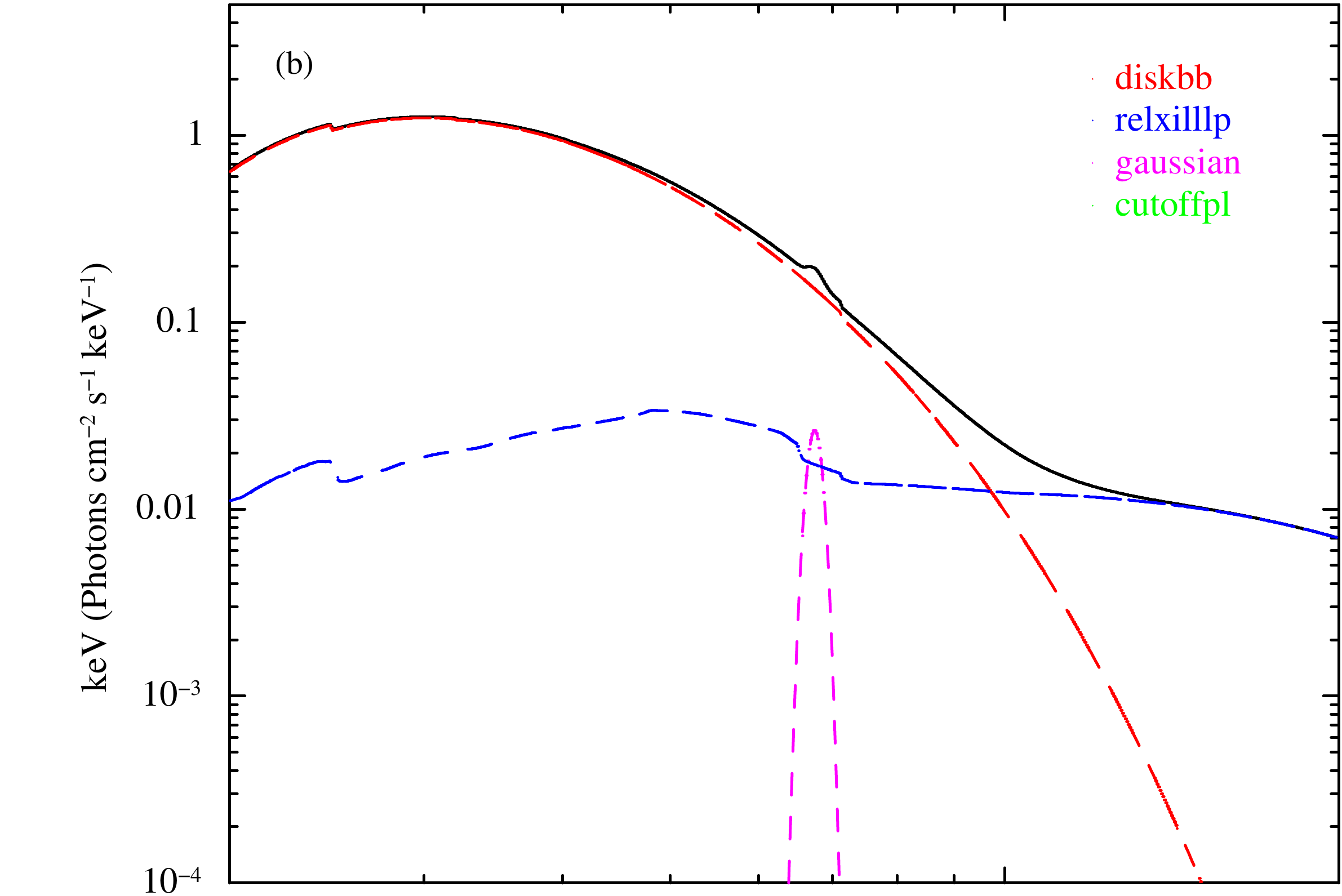}} 
    \mbox{\includegraphics[angle=270,width=0.9\linewidth]{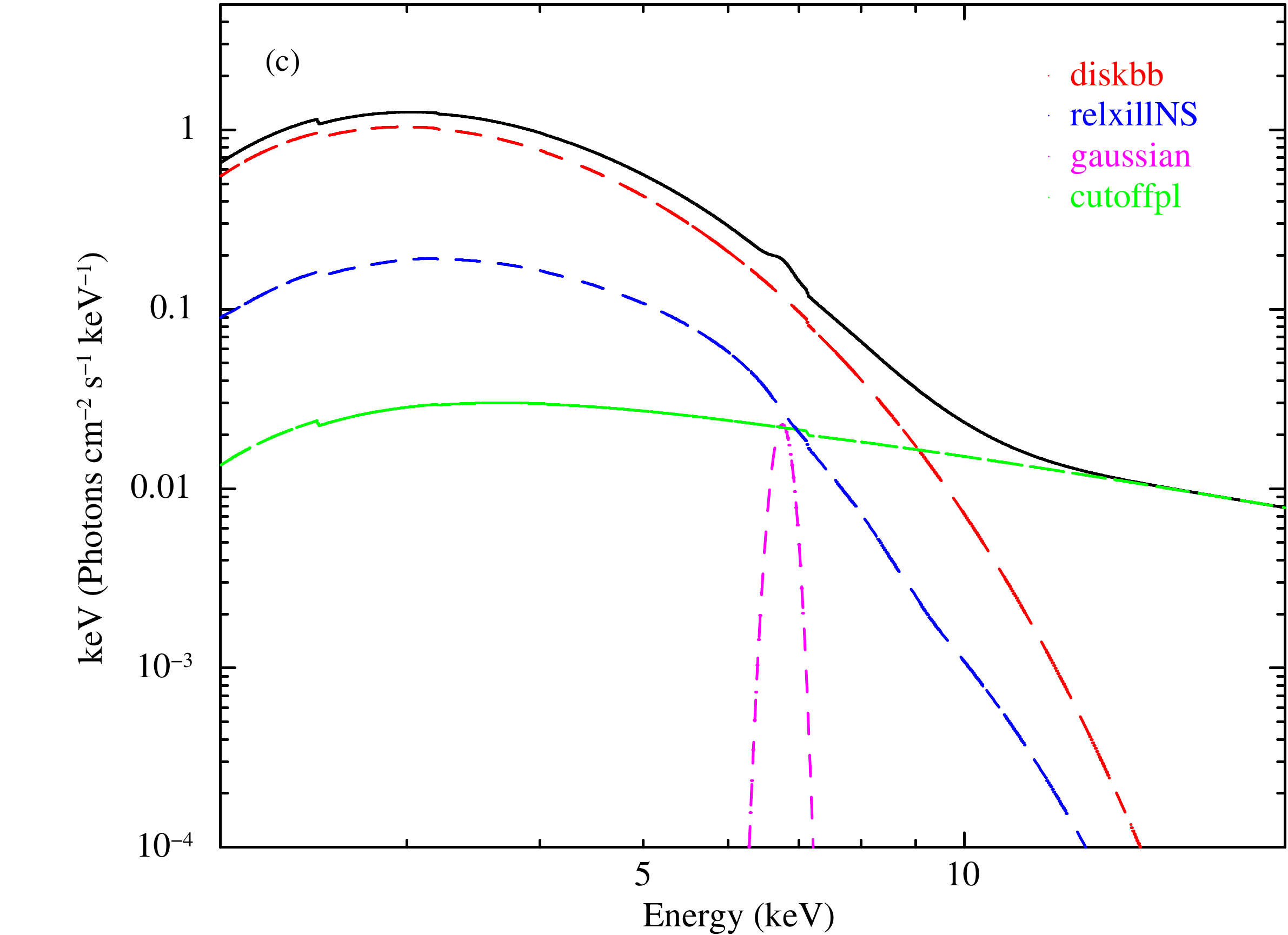}}
    
    \caption{Individual components in the fit to the {\hx} spectra of {\exo} in the 2--20~keV. The component \texttt{cutoffpl} in the first two figures is negligible.}
    \label{fig:mo_hxmt}
\end{figure}

Both features, an emission line at $\sim6.7$~keV and an excess at above 10~keV, hint to the presence of a reflection spectrum.
After adding the reflection component \texttt{relxill} to the model, the fit improved with $\Delta \chi^2=76.9$ for 6 dof fewer. However, Fig.~\ref{fig:ra_hxmt}b shows that the emission feature still appears. We hence added a \texttt{gaussian} component to account for this feature which improves the spectral fit by $\Delta \chi^2 = 40.5$ for 3 dof fewer.
The overall model is \texttt{tbabs*(diskbb+gaussian+relxill+cutoffpl)}. Here \texttt{relxill} only accounts for the reflection spectrum when a negative reflection fraction is adopted. We show the individual component in Fig.~\ref{fig:mo_hxmt}a and the best-fitting parameters, as well as the individual flux of each component, in Table~\ref{tab:hxmt}.

The disk temperature obtained from the fit to the {\hx} spectra is $0.94\pm0.01$~keV. The inner radius of the accretion disk is between $2.7-4.1~R_{\rm g}$. Compared to the intermediate state, the higher \texttt{diskbb} temperature, higher photon index and smaller inner radius are consistent with the fact that the source evolves to the soft state. However, the hard component in this model is somehow negligible: adding this component has no effect on the fit. This result seems very odd that a reflection spectrum exists without an illuminating source, i.e. a hard coronal component in this case.

We then replaced the reflection component \texttt{relxill} with its lamppost version \texttt{relxilllp} to fit the {\hx} spectra. Again, the narrow emission line at $\sim6.7$~keV appears on top of the broad line (see Fig.~\ref{fig:ra_hxmt}c). After adding a \texttt{gaussian} component, the best-fitting parameters in both models are consistent (see the relevant results in Figs.~\ref{fig:ra_hxmt} and \ref{fig:mo_hxmt} and Table~\ref{tab:hxmt}). Similar to the result with \texttt{relxill}, the hard component in \texttt{relxilllp} is also negligible.

We use the ratio, $R_{\rm s}$, to parameterize the strength of the reflection as a ratio of the fluxes of the reflection component to that of the incident continuum in a selected energy band, as used by e.g.  \cite{Tao2015,Dauser2016}. We calculate $R_{\rm s}$ in the 10--20~keV: for the case of \texttt{relxill}, $R_{\rm s} \approx 2\times10^8$ and for the case of \texttt{relxilllp}, $R_{\rm s} \approx 5\times10^8$. \cite{Dauser2016} report that the reflection strength is strongly dependent on the disk inclination angle and the coronal height if a lamppost geometry is assumed. Nevertheless, for no configuration such a high reflection strength is expected.

Inspired by the work of \cite{Connors2020}, a reflection spectrum can be produced by the disk self-irradiation when the source is in a very soft state. While the inner accretion disk approaches the inner stable circular orbit (ISCO), the gravitational pull of the black hole could bend the thermal photons back and illuminate itself. 
The model, \texttt{relxillNS} (T.~Dauser \& J.~{Garc{\'\i}a}, in prep), was originally designed for the reflection originating from the accretion disk in a neutron-star system, in which the illuminating source is a single-temperature blackbody. As in \cite{Connors2020}, we fixed the inner radius in \texttt{relxillNS} at the ISCO where the \texttt{diskbb} temperature is assumed to be sufficient enough to illuminate the disk. 
The electron density and the iron abundance of the accretion disk are degenerated and always pegs at their upper limits; we thus fixed the electron density, $\log~(n\rm_{e}/cm^{-3})=19$, and allowed the iron abundance to vary. 

We show the result of the fit with \texttt{relxillNS} in Figs.~\ref{fig:ra_hxmt} and \ref{fig:mo_hxmt} and Table~\ref{tab:hxmt}. The inclination angle and the iron abundance are consistent with the values obtained in the intermediate state. Besides that, the photon index derived from \texttt{relxillNS} is $1.97\pm0.09$, much lower than the ones derived from \texttt{relxill} and \texttt{relxilllp}.

If we do not include any reflection component in the model (see the residuals in Fig.~\ref{fig:ra_hxmt}f), the fit becomes significantly worse, $\Delta \chi^2=75.1$ for 7 dof more, with associated F-test probability $\sim1\times 10^{-13}$. 
In this step. we fixed the centroid energy and the width of \texttt{gaussian} at their best-fitting values, i.e. $E_{\rm gau}=6.73$~keV and $\sigma_{\rm gau}=0.14$~keV, to avoid affecting other features.
Fig.~\ref{fig:ra_hxmt}f shows that besides the high energy hump, there is a slight excess centered at $5-6$~keV. Adding a single blackbody component, \texttt{bbody} in {\sc xspec}, we obtained a fit with $\chi^2=965.2$ for 988 dof which is worse than the fit with \texttt{relxillNS} with $\Delta\chi^2=21.2$ for 5 dof more. Meanwhile, the best-fitting photon index increases to $5.56_{-0.53}^{+1.71}$ which makes this case less convincing.

Figs.~\ref{fig:ra_hxmt}b, c and d show that the residual feature at $\sim6.7$~keV is independent of the choice of the reflection component. 
We calculated the line significance by simulating 1000 spectra using the {\sc xspec} script simftest in which the centroid energy of the \texttt{gaussian} component are allowed to vary within its 1-$\sigma$ confidence region.
Based on the model with \texttt{relxillNS}, the observed change in $\chi^2$ is 37.9 and the probability of finding such a change by chance is $1.1 \times 10^{-21}$ which corresponds to 9.5$\sigma$ significance. 

To better understand the nature of this narrow emission line, we tested two possibilities here. We first tried the possibility of a distant reflector, fitting the line with an unblurred reflection component \texttt{xillver}. We linked all the parameters in \texttt{xillver} to those in \texttt{relxillNS} except for the ionization parameter and normalization. 
Neither setting the ionization parameter in \texttt{xillver} to be zero nor linking it to that in \texttt{relxillNS} could fit the line. The fit to the spectra with the model \texttt{tbabs*(diskbb+xillver+relxillNS+cutoffpl)} is reasonable, $\chi^2=946.0$ for 984 dof. However, the best-fitting ionization parameter is $3.77_{-0.02}^{+0.17}$, even higher than the one derived from \texttt{relxillNS}. This result is in conflict with our initial assumption of a distant reflection. 


We then tried to see if the narrow line is the re-emission from disk winds although without the sign of an absorption feature. We produced a grid of models with {\sc xstar} again but adopting a \texttt{bbody} input spectrum with a blackbody temperature of 0.92~keV. 
We applied the same assumptions for the gas density and covering fraction as listed in Table~\ref{tab:xstar} but changed the luminosity and the turbulent velocity to $3.9 \times 10^{38}~\rm erg~s^{-1}$ and 5000~$\rm km~s^{-1}$, respectively, based on the best-fitting parameters for the {\hx} spectra.
We replaced the component \texttt{gaussian} with the precalculated grid and obtained the best fit with $\chi^2=942.1$ for 982 dof, the hydrogen equivalent column density, $10^{15}<N< 10^{22}$, the ionization parameter, $\log \xi=4.41_{-0.44}^{+0.24}$, the blueshifted velocity, $v<0.008c$ and the normalization, $N=0.13_{-0.08}^{+66.44}$. Apparently our data cannot constrain this complex photoionization model.

\begin{table*}
\caption{Best-fitting parameters of the {\hx} spectra of {\exo} with the models \texttt{tbabs*(diskbb+gaussian+relxill+cutoffpl)}, \texttt{tbabs*(diskbb+gaussian+relxilllp+cutoffpl)} and \texttt{tbabs*(diskbb+gaussian+relxillNS+cutoffpl)} in the energy band 2--9~keV and 10--20~keV.}
\renewcommand{\arraystretch}{1.4}
\setlength{\tabcolsep}{2pt}
\footnotesize
\centering
\begin{tabular}{clcccccc}
\hline \hline
\multicolumn{2}{c}{Components }&\multicolumn{3}{c}{{\em Insight-HXMT}}  \\
\texttt{tbabs} &$N\rm_{H}~(10^{22}~cm^{-2})$ &$6.1\pm0.1$&$6.1\pm0.1$ &$6.0\pm0.1$\\
\hline
\texttt{diskbb}&$T\rm_{dbb}$~(keV)& $0.94\pm0.01$ &$0.94\pm0.01$ & $0.92\pm0.01$\\
&$N\rm_{dbb}$~$(R_{\rm dbb}^2/D_{10}^2 \cos~i)$&$1436.1_{-52.5}^{+26.2}$ &$1439.6_{-26.6}^{+1.1}$ & $1294.9_{-205.4}^{+159.6}$\\
\noalign{\smallskip}
\hline                              
\texttt{relxill(lp/NS)}& $q\rm_{in}$ &$4.8\pm0.8$ & - &$3.9_{-0.3}^{+0.9}$\\
& $h~(R\rm_{g})$ &-  &$<2$ &-\\
& $\log n~(\rm cm^{-3})$ &-  &- & $19^f$\\
& $A_{\rm Fe}~(A_{\rm Fe_{\odot}})$ &$1.5_{-0.5}^{+1.0}$&$1.2\pm0.2$&$4.9_{-1.9}^{10p}$\\
& $i$~($^{\circ}$)& $26.6_{-13.1}^{+5.6}$&$17.0_{-9.5}^{+6.0}$&$<36.5$\\
& $a_{*}$ &$0.998^{f}$&$0.998^{f}$&$0.998^{f}$\\
&$E_{\rm cut}$~(keV) &$500^f$ &$500^f$ &$500^f$ \\
&Refl\_frac&$-1^{f}$ &$-1^{f}$ &$-1^{f}$     \\
&$R_{\rm in}$&$3.4\pm0.7~R_{\rm g}$& $3.9_{-0.1}^{+0.9}~R_{\rm g}$ & $1^f~R_{\rm ISCO}$\\
&$\log \xi$~($\rm erg~cm~s^{-1}$)&$2.28_{-0.54}^{+0.05}$ &$2.10_{-0.04}^{+0.19}$ &$2.11_{-0.10}^{+0.39}$ \\
&$N_{\rm rel}/N_{\rm relp}/N_{\rm relns}$&$0.06\pm0.03$ &$9.7_{-0.1}^{+10.1}$&$0.009\pm0.003$\\ 
\noalign{\smallskip}
\hline  
\texttt{cutoffpl}&$\Gamma$&$3.02_{-0.04}^{+0.09}$ &$3.10\pm0.12$ & $1.97\pm0.09$\\
&$N_{\rm pl}$&$<0.14$&$<0.3$&$0.15\pm0.05$&\\
\hline                              
\texttt{gaussian}&$E_{\rm gau}$~(keV)&$6.73\pm0.03$&$6.74\pm0.02$&$6.75\pm0.03$\\
&$\sigma\rm_{gau}$~(keV)&$0.14\pm0.04$&$0.11\pm0.04$&$0.14\pm0.05$\\
&strength ($10^{-3}$)&$1.3\pm0.2$ &$1.1\pm0.2$ &$1.3\pm0.4$\\
\hline
\texttt{const}&LE&$1^f$&$1^f$&$1^f$\\
&ME&$0.99\pm0.03$&$0.98\pm0.02$&$0.93\pm0.03$\\
\hline
\multicolumn{2}{c}{$\chi^{2}$/dof} &941.2/982 & 944.1/981 & 944.0/983\\
\hline
&$F\rm_{dbb}$&$10.1\pm0.1$ &$10.1\pm0.1$&$8.4_{-1.5}^{+0.9}$\\
&$F\rm_{rel}/F\rm_{relp}/F\rm_{relns}$&$0.4\pm0.1$ &$0.4\pm0.1$&$1.6_{-0.8}^{+1.5}$ \\
&$F\rm_{pl}$&$<0.09$&$<0.16$&$0.4\pm0.1$ \\
&$F\rm_{tt}$&$10.5\pm0.1$ &$10.5\pm0.1$&$10.4\pm0.1$\\
\hline
\end{tabular}
\begin{flushleft}
{\bf Note:} The symbols $p$ means that the parameter pegs at its limit; $f$ means that the parameter is fixed during the fit. All the unabsorbed fluxes, $F\rm_{dbb}$, $F\rm_{rel}$, $F\rm_{relp}$, $F\rm_{relns}$, $F\rm_{pl}$ and $F\rm_{tt}$, are calculated in the energy band of 2$-$10~keV in units of $10^{-9}~\rm erg~cm^{-2}s^{-1}$. 
\end{flushleft}    \label{tab:hxmt}
\end{table*}

\section{discussion}\label{sec:discussion}
After over three decades of quiescence, the black-hole candidate {\exo} was again active. We studied its spectral properties in the hard intermediate and soft states in the 2019 outburst with {\Nu} and {\hx} observations. 
A reflection component is present in both states that becomes weaker from the intermediate to the soft state. Fits with a reflection component yield an inclination angle of the accretion disk of $\sim40\degree$ with respect to the line of sight. Besides that, a moderately broad absorption line has been observed at $\sim7.2$~keV in the intermediate state but it disappears in the soft state, whereas a narrow emission is present at $\sim6.7$~keV instead. In the following sections we discuss the possible origin of the absorption and emission lines observed in {\exo}.

\subsection{Disk wind in the hard intermediate state}\label{sec:disk_wind}
There is evidence for an anti-correlation between jets and disk winds in LMXBs: jets are present in the hard state and disk winds are present in the soft state (e.g. \citealt{Miller2006b,Miller2008}). However, some studies show that jets and disk winds can co-exist in the hard state (e.g. \citealt{Homan2016,Gatuzz2020}). Moreover, \cite{Homan2016} suggested that disk winds and jets do not need to be mutually exclusive in LMXBs when a source luminosity is higher than a few tens of percent of the Eddington luminosity.

Radio emissions of {\exo} have been detected with the VLA on August 1 \citep{MillerJones2019} and MeerKAT on August 4 \citep{Williams2019}, respectively, indicating the presence of a jet. The {\Nu} observation, accompanied with an absorption line at $\sim 7.2~\rm keV$ was performed on August 3, in between the radio detection. Therefore, the jet and disk wind are probably present simultaneously in {\exo}. Adopting a distance of 8~kpc to the source \citep{MillerJones2019} and a BH mass of 3.24~$M_{\odot}$ \citep{Strohmayer2020} gives an X-ray luminosity of $\sim0.1~L_{\rm EDD}$ on August 3, consistent with the conclusion reported by \cite{Homan2016}. 
Adopting a larger BH mass, the corresponding X-ray luminosity would be lower than $\sim0.1~L_{\rm EDD}$, suggesting that for the co-existence of jets and disk winds, it is not essential that a source is brighter than tens of percent of the Eddington luminosity.

Thermally driven winds may occur when the launching radius is greater than the Compton radius, $R_{\rm C}$, where the gravitational and thermal pressures are equal. Previous work suggests that such winds can even be driven from $0.1~R_{\rm C}$ \citep{Begelman1983,Woods1996}.
The Compton radius is given by

\begin{equation} 
R_{\rm C}= \frac{10^{10}}{T_{\rm C,8}}\times \frac{M_{\rm BH}}{M_{\odot}} \rm~cm,\\
\label{equa1}
\end{equation}
where $T_{\rm C,8}$ is in units of $10^8$~K. 
We calculate the Compton temperature via 
\begin{equation} 
T_{\rm C,8}= \frac{h \int \nu E_{\rm \nu } d\nu }{4k \int E_{\rm \nu } d\nu },\\
\label{equa2}
\end{equation}
where $E_{\rm \nu}$ is the illuminating spectrum, $h$ is the Planck constant and $k$ is the Boltzmann constant \citep{Frank1985}.
When the spectrum is dominated by a cutoff power law in the intermediate state, we assume $E_{\rm \nu }=Ke^{-\frac{h\nu}{E_{\rm cut}}}(h\nu)^{-\Gamma}$, where $K$ is the normalization of \texttt{cutoffpl}. 
Adopting the best-fitting photon index and cutoff energy for either the low- or high-flux segment, this gives a Compton temperature of $T_{C,8}=0.11\pm0.01$. Bringing this Compton temperature to Eq.~\ref{equa1}, we obtain $R_{\rm C}=29.5\pm0.1 \times 10^{10}$~cm when adopting a BH mass of 3.24~$M_{\odot}$ from \cite{Strohmayer2020}.

The launching radius is therefore about three orders of magnitude smaller than the Compton radius, $R_{\rm launch}=0.001~R_{\rm C}$, for both segments. If we considered a larger BH mass for {\exo}, the Compton radius would increase accordingly and hence the launching radius would be smaller, making it even less likely that such wind is driven by thermal expansion.
Although by taking radiation pressure due to electron scattering into account, the launching radius could extend down to $0.01~R_{\rm C}$ (still one order of magnitude larger than our measurement), this requires a near-Eddington luminosity \citep{Proga2002} or a sub-Eddington luminosity when the gas temperature $< 10^5$~K \citep{Proga2007}.
As to line interaction, it is only effective for the gas with low ionization \citep{Proga2000}. We therefore rule out the possibility of a radiation-driven wind in our case. This leaves magnetic forces as the only viable driving mechanism for the observed wind in {\exo}. 

Although disk winds have been largely detected in high-inclination systems, they have been detected in some low-inclination systems as well, for instance GX~340+0 \citep{Miller2016}, 1RXS~J180408.9-342058 \citep{Degenaar2016} and MAXI~J1631-479 \citep{Xu2020}. Thermally driven winds are concentrated in the plane of the disk and hence the line of sight is more likely to intercept the wind in high-inclination systems (e.g. \citealt{Miller2006b,Tomaru2020}). On the other hand, \cite{Chakravorty2016} suggests that magnetically driven winds can be observable at low inclination angles, which would match with our observations, although \cite{Chakravorty2016} also reports that such winds can only be produced in the soft states. 
In conclusion, further studies are required to understand the observed winds/absorption lines in {\exo}.

\cite{Draghis2020} applied the same reflection models \texttt{relxill(lp)} to fit the same {\Nu} observation of {\exo} as we did in this work. With a different model assumption, they derived different constraints of the system properties, e.g. a high spin parameter, $a_{*}=0.995-0.998$, a high inclination angle of the accretion disk, $i \approx 73\degree$ and a sub-solar iron abundance, $A_{\rm Fe}=0.58-0.87$. 
We re-examined the {\Nu} data to explore why we obtained so divergent results with the model \texttt{tbabs*(diskbb+relxill)}. The absorption line model is removed to avoid the effect of the degeneracy between \texttt{relxill} and \texttt{gabs}. The result shows that the reflection fraction ($\rm refl\_frac$ in \texttt{relxill}) played a key role here. If we freeze the refection fraction to be 1, as \cite{Draghis2020} assumed in their work, let the spin parameter free, and keep the other assumption for \texttt{relxill} as described in Section~\ref{sec:nu} in the fit, we would obtain a spin parameter, $a_{*}=0.96\pm0.02$, a much higher inclination angle, $i=84.4\pm0.2\degree$ and a sub-solar iron abundance, $A_{\rm Fe}=0.67\pm0.03$, with a worse fit $\Delta\chi^2=112.3$ for 1 dof more. Because we split the {\Nu} observation into two segments, if we use the same assumption as in \cite{Draghis2020}, we would have more free parameters in the model and thus cannot constrain parameters as well as they did.
Since the only thing we want to show here is that the best-fitting parameters \cite{Draghis2020} obtained are strongly dependent on the value of the reflection fraction, we did not conduct this test further.

\subsection{Illuminating source in the soft state}
We conducted a spectral analysis of the observations of {\exo} in both the intermediate and soft states. For the spectra in the intermediate state, we fitted the continuum with the same model but fitted the reflection spectrum with either \texttt{relxill} or \texttt{relxilllp}. The best-fitting parameters are consistent within the two models and are in good agreement with the spectral state. However, the fit to the spectra in the soft state is debatable. As described in Section~\ref{sec:hx}, we have fitted the {\hx} spectra with three versions of the reflection models, \texttt{relxill}, \texttt{relxilllp} and \texttt{relxillNS}, which assume different types and geometries of its illuminating source. 

In the fits with either \texttt{relxill} or \texttt{relxilllp}, the hard component, which is the required illuminating source for both models, is negligible. We define the reflection strength as a ratio of the reflected to the incident flux in the 10--20~keV where we see a Compton hump in the {\hx} observation. The reflection strength is $2\times10^8$ for \texttt{relxill} and $5\times10^8$ for \texttt{relxilllp}, both of which are unrealistically high for any configuration.

Moreover, the best-fitting photon index derived from \texttt{relxill} or \texttt{relxilllp} is around 3. Such a high value of the photon index is normally observed in a `very high' state (VHS) or  `steep power-law' (SPL) state in which the spectrum is characterized by a strong power-law component \citep{Miyamoto1993,Klis2004}. This is in conflict with the fact that the soft component \texttt{diskbb} contributes over 96\% of the total flux in this observation when fitting the reflection spectrum with either \texttt{relxill} or \texttt{relxilllp}. If we freeze $\Gamma$ at 2, the fit worsens by $\Delta \chi^2=24.6$ meanwhile the best-fitting inclination angle pegs at $3\degree$, not consistent with the value derived from the fit to the {\Nu} spectra. Overall, both models are very doubtful for explaining the reflection features present in the {\hx} observation.

In terms of \texttt{relxillNS}, the model fits the data well and the derived iron abundance and the inclination angle of the accretion disk are in accordance with the ones derived from the fit to the {\Nu} spectrum in the intermediate state. 
These results leaves us a possibility that we may observe another case of the disk self-irradiating reflection spectrum in a BHXB (the first one is reported by \citealt{Connors2020}). When the inner radius of the accretion disk reaches the ISCO, the gravity of the central compact source is strong enough to bend the light from the disk on itself. 
To explore if this is a physically viable theory for {\exo}, we calculate the fraction of photons returning to the disk based on General relativistic ray tracing simulations (see more details in \citealt{Connors2020}). Using the code developed by \cite{Yang2013} and adopting a standard prescription for the disk emission with zero torque condition \citep{Shakura1973,Poutanen2007}, we estimated the fraction as a function of spin. Assuming a spin of 0.998, the estimated fraction is about 18\%, in good agreement with the observed value, i.e. $F_{\rm relns}/F_{\rm tt}=18\%$ (see Table~\ref{tab:hxmt}). This result suggests that disk self-irradiation could be the origin of the observed reflection feature in {\exo}.
However, it should be kept in mind that this is just a rough estimate of the photons returning to the disk, which could be very different from what the disk sees in reality when taking General Relativity energy shifts into account. Further exploration is beyond the scope of this work. More details will be presented in T.~Dauser \& J.~{Garc{\'\i}a} (in prep).

\subsection{Origin of the narrow emission line in the soft state}
Disk winds, identified by absorption lines have been commonly observed in LMXBs in the soft state (e.g. \citealt{Miller2008,Neilsen2009,Miller2015}). However, we only observed absorption lines in the intermediate rather than the soft state in {\exo}. 
As we mentioned in Section~\ref{sec:hx}, a narrow emission line at $\sim6.7$~keV is present on top of the reflection component regardless of the choice of a reflection model. We have tested two possibilities for its origin, a distant reflection or re-emission from disk winds. The former has been ruled out since, for the same electron density, the fit of a distant reflection yields an even higher ionization parameter than the near one. The latter cannot be confirmed either since the overall model is too complex to be constrained by the {\hx} data. 

\cite{Diaz2014} suggested another possible origin for a narrow emission line present in BHXBs. By analyzing six {\em XMM-Newton} observations of 4U~1630--47, \cite{Diaz2014} interpreted the disappearance of a disk wind, i.e. absence of an absorption line, as a consequence of strong ionization of the wind when 4U~1630--47 transited from the high soft to the VHS/SPL and attributed a narrow emission line above 7~keV to arise in an optically thin jet while 4U~1630--46 was in the VHS/SPL. 
Although without including any radio observation in our work it is hard to tell the presence of jets in {\exo} in the soft state, the observed hard emission seems too weak to be responsible for that. 

If the narrow line is not from disk winds, then the disk winds observed in the intermediate state disappeared in the soft state. Since the input source in the soft state is much brighter than the one in the intermediate state, the disk wind could become `transparent' due to a strong ionization as \cite{Diaz2014} suggested. Overall, observations with the higher spectral resolution are required to examine this result.

\section{conclusion}
We analyzed {\Nu} and {\hx} observations of {\exo} in the hard intermediate and soft states. By fitting a reflection component we obtained an inclination angle of $\sim40\degree$ with respect to the line of sight. Our fit of the absorption line at $\sim7.2$~keV indicates a highly ionized fast outflowing disk wind with an ionization parameter, $\log~\xi>6.1$, and a velocity up to 0.06c. The quasi-simultaneous radio emissions reported in the literature suggest jets and disk winds probably co-exist in {\exo}.
With the modest spectral resolution of {\Nu}, the observed wind appears to be magnetic driven. 

A weak reflection component has been detected in the {\hx} observation in the soft state. We tried to fit it with the different versions of reflection models \texttt{relxillNS}, \texttt{relxill}, and \texttt{relxilllp} and all of them returned a reasonable fit. For the fit with \texttt{relxillNS}, this model describes the reflection spectrum well and the obtained inclination angle and the iron abundance are consistent with the ones derived in the intermediate state, implying a possible detection of self-irradiating disk reflection.
Assuming a standard accretion disk with zero torque condition \citep{Shakura1973}, \cite{Connors2020} found that $\sim5\%$ of the photons could return to the disk when a spin of 0.5 is adopted, which is well matched with their observation. 
In the case of {\exo}, the fraction of the reflected flux with respect to the direct emission is $\sim18\%$, over three times larger than their detection. However, using the same assumption and meanwhile adopting a spin of 0.998, the estimation of the returning photons is also being three times larger than that in \cite{Connors2020}. Overall, this result states that disk self-irradiation is able to produce the reflection features observed in {\exo} in the soft state. As to the latter two models, both of them require a very large photon index and an extremely high reflection strength, making both scenarios less convincing. The origin of the narrow emission line at $\sim6.7$~keV is still debatable. X-ray data with the higher spectral resolution, e.g. {\em Chandra} and {\em X-Ray Imaging and Spectroscopy Mission} ({\em XRISM}; \citealt{Tashiro2018,XRISM2020}), are required to further investigate this.

\acknowledgments
{We are grateful to Timothy R.~Kallman for help with {\sc xstar}. We thank the anonymous referee for the helpful comments.
This work made use of the data from the {\em Insight-HXMT} mission, a project funded by China National Space Administration (CNSA) and the Chinese Academy of Sciences (CAS). The {\em Insight-HXMT} team gratefully acknowledges the support from the National Program on Key Research and Development Project (Grant No. 2016YFA0400800) from the Minister of Science and Technology of China (MOST) and the Strategic Priority Research Program of the Chinese Academy of Sciences (Grant No. XDB23040400). The authors thank supports from the National Natural Science Foundation of China under Grants No. 11673023, 11733009, 11603037, 11973052, U1838201, U1838115, U1938103 and U1838202. 
Y.~W. and L.~Z. acknowledge support from the Royal
Society Newton Funds. DA acknowledges support from the Royal Society. 
J.~M. acknowledges the support from STFC (UK) through the University of Strathclyde UK APAP network grant ST/R000743/1.
J.A.G. acknowledges support from NASA grant 80NSSC20K1238 and from the Alexander von Humboldt Foundation. Additionally, this work has made use of data from the NuSTAR mission, a project led by the California Institute of Technology, managed by the Jet Propulsion Laboratory, and funded by the National Aeronautics and Space Administration. We thank the NuSTAR Operations, Software, and Calibration teams for support with the execution and analysis of these observations. This research has made use of the NuSTAR Data Analysis Software (NuSTARDAS), jointly developed by the ASI Science Data Center (ASDC, Italy) and the California Institute of Technology (USA).}
\software{XSPEC (v12.11.1; \citealt{Arnaud1996}), NuSTARDAS (v2.0.0), HXMTDAS (v2.02; \citealt{Zhang2020}).} \\

\bibliographystyle{aasjournal}
\bibliography{sample63}

\end{document}